\documentclass{aa}
\pdfoutput=1 
\usepackage{appendix}
\usepackage{latexsym}
\usepackage{graphicx} 
\usepackage{lscape}
\usepackage{afterpage}
\usepackage{booktabs}
\usepackage{multirow}
\usepackage{arydshln}
\usepackage{ulem}

\usepackage{ulem}
\usepackage[citecolor =blue]{hyperref}
\usepackage{psfrag,color,ulem,gensymb}
\definecolor{deepblue}{rgb}{0,0,0.5}
\definecolor{deepred}{rgb}{0.6,0,0}
\definecolor{deepgreen}{rgb}{0,0.5,0}
\definecolor{darkgreen}{rgb}{0,0.6,0}
\definecolor{darkspringgreen}{rgb}{0.09, 0.45, 0.27}
\definecolor{debianred}{rgb}{0.84, 0.04, 0.33}
\newcommand{\kms}{km s$^{-1}$~}
\newcommand{\kmsb}{km s$^{-1}$}
\newcommand{\pa}{Pa$\alpha~$}

\newcommand{\brd}{Br$\delta$}

\newcommand{\hmol}{H$_2$~}

\newcommand{\msun}{M$_{\rm \odot}$~}
\newcommand{\msunb}{M$_{\rm \odot}$}
\newcommand{\ergcms}{erg s$^{-1}$ cm$^{-2}$}
\newcommand{\ergs}{erg s$^{-1}$}

\usepackage{amstext}

\begin{document}

\title{Limited impact of jet induced feedback in the multi-phase nuclear interstellar medium of   4C12.50}
\author{M. Villar Mart\'{i}n$^{1}$, N. Castro-Rodr\'\i guez$^{2,3}$, M. Pereira Santaella$^4$, I. Lamperti$^1$,  C. Tadhunter$^5$, B. Emonts$^6$, L. Colina$^{1}$, A. Alonso Herrero$^{7}$, A.Cabrera-Lavers$^{2,3}$, E. Bellocchi$^{8,9}$}
\institute{$^{1}$Centro de Astrobiolog\'{i}a (CAB), CSIC-INTA, Ctra. de  Ajalvir, km 4, 28850 Torrej\'{o}n de Ardoz, Madrid, Spain\\
$^2$GRANTECAN, Cuesta de San Jos\'e s/n, E-38712, Bre\~na Baja, La Palma, Spain \\
$^3$Instituto de Astrof\'\i sica de Canarias, V\'\i a L\'actea s/n, E-38200 La Laguna, Tenerife, Spain \\
$^4$Observatorio Astron\'omico Nacional (OAN-IGN)-Observatorio de Madrid, Alfonso XII, 3, 28014 Madrid, Spain
 \\
$^5$Department of Physics \& Astronomy,    The Hicks Building, University of Sheffield, Sheffield S3 7RH, UK \\
$^6$National Radio Astronomy Observatory, 520 Edgemont Road, Charlottesville, VA 22903, USA \\
$^7$Centro de Astrobiolog\'\i a (CAB), CSIC-INTA, Camino Bajo del Castillo s/n, 28692 Villanueva de la Ca\~nada, Madrid, Spain \\
$^8$Departamento de F\'\i sica de la Tierra y Astrof\'\i sica, Fac. de CC F\'\i sicas, Universidad Complutense de Madrid, 28040 Madrid, Spain\\
$^9$   Instituto de F\'\i sica de Part\'\i culas y del Cosmos IPARCOS, Fac. CC F\'\i sicas, Universidad Complutense de Madrid, 28040 Madrid, Spain
 \\
 \email{villarmm@cab.inta-csic.es}}        
\date{}
\abstract 
{4C12.50 (IRAS 13451+1232) at $z=$0.122  is  an ultraluminous infared radio galaxy that has often been proposed as a prime candidate for the link between ultraluminous infared galaxies and young radio galaxies. It is also an interesting target to investigate whether and how radio induced feedback may affect the evolution of galaxies in the early phases of radio activity.}
{We study in detail for the first time the hot ($\ge$1500 K) molecular gas  in 4C12.50.  The potential impact of the radio jet on this gas phase, as well as on the star formation activity, are  investigated. We also study the ionized (including coronal) gas as traced by the near-infrared lines.}
{Using near-infrared long slit spectroscopy obtained with EMIR on GTC and Xshooter on VLT, we analyse the emission line spectrum of the ionized, coronal and, specially, the hot molecular gas in the western nucleus hosting the compact radio jet. Based on high spatial resolution ALMA CO(2-1) data, we also revise the location of 4C12.50 in the Kennicutt-Schmidt diagram in order to investigate whether star formation is suppressed.}
{4C12.50 hosts  (2.1$\pm$0.4)$\times$10$^{4}$ M$_{\rm \odot}$ of hot molecular gas. An unusually high rotational temperature $T_{\rm rot}$ =3020$\pm$160 K is inferred. The molecular gas mass obeys  a power law temperature distribution, $\frac{d M_{\rm H2}}{dT}\propto T^{-5}$  from T$\sim$ 300 K and up to  $\sim$3000 K.  Both results support that shocks (probably induced by the radio jet) contribute to the heating and excitation of the hot molecular gas.  A  molecular outflow is not detected. The  coupling of the outflowing ionized and neutral outflows with the  hot molecular gas is poor. Contrary to other studies, we claim that there is  no evidence for star formation suppresion in this object.}
{If radio induced feedback can regulate the star formation activity in galaxies, 4C12.50 is a promising candidate to reveal this phenomenon in action.  However, we find   no solid evidence for current or past impact of this mechanism on the evolution of this system, neither by clearing out the dusty central cocoon efficiently, nor by suppressing the star formation activity.}


\titlerunning{Jet induced feeback in 4C12.50.}
\authorrunning{Villar-Mart\'\i n et al.}

\maketitle

\section{Introduction}
\label{intro}

Gigahertz-Peaked Spectrum (GPS) sources are compact and often powerful radio sources. They  are estimated to  make up around 10\%   of the bright radio-source population (O'Dea et al. \citeyear{ODea1998,ODea2021}, \citealt{Sadler2016}). There are three main hypotheses for their nature, which can vary from source to source. They might be: (1)   very young radio galaxies which will evolve into  large radio galaxies. (2) compact due to the confinement  by interactions with dense gas in their environments;  (3)  transient or intermittent sources (\citealt{ODea2021}).

The GPS sources are entirely contained within the extent of the narrow-line region (NLR, $\la$1 kpc). 
Because of the similar spatial scales, the feedback effects that result from the interaction between the radio source and the dense circumnuclear interstellar medium (ISM) can be very strong. Thus, they are interesting targets to investigate whether and how radio induced feedback may affect the evolution of galaxies in the early phases of radio activity (\citealt{ODea1998,Holt2003,Holt2011,Morganti2013,Santoro2020}).

 4C12.50 (IRAS 13451+1232) at $z=$0.122  (luminosity distance $D_{\rm L}$=573 Mpc) is one of the closest and best known GPS sources ($P_{\rm 5GHz}\sim$10$^{33}$ W Hz$^{-1}$; \citealt{ODea1998,Holt2003}).  Radio emission stretching  outside the host galaxy provides evidence of a previous radio outburst, that occured $\sim10^{7-8}$ yr ago (\citealt{Stanghellini2005}). The jet may have restarted only recently  ($<$10$^5$ yr) after a long period of inactivity or be a central component in a  continuous supply of energy from the  core to the extended lobes (\citealt{Lister2003,Odea2000,Stanghellini2005,Morganti2013}).
Jet frustration appears to be working at some level, but the amount of mass seems to be insufficient to confine the jet completely (\citealt{Morganti2004,Morganti2013}).

4C12.50 has often been suggested to be a prime candidate for the link between ultraluminous infrared galaxies
 (ULIRGs) and young radio galaxies (\citealt{Gilmore1986}, Morganti et al.   \citeyear{Morganti2003,Morganti2013}).  
It is  a ULIRG with $L_{\rm IR}$ =log($L_{\rm 8-1000 \mu m}$/L$_{\rm \odot}$)=12.31 and a  star forming rate SFR$\sim$100 \msun yr$^{-1}$ (\citealt{Mirabel1989b,Rupke2005a,Perna2021,Pereira2021}).
4C12.50 is hosted by an elliptical galaxy with two optical nuclei separated by 1.8$\arcsec$ or 4.0 kpc.  Additional morphological signs reveal a major merger event  which is in the later stages involving at least one gas-rich galaxy (e.g. \citealt{Heckman1986,Emonts2016}). The western, primary nucleus is active and hosts the compact jet. This has a small  size ($\sim$220 pc) and twin-jet morphology. (\citealt{Grandi1977,Gilmore1986,Veilleux1997}).  The high [OIII]$\lambda$5007 luminosity $L_ {\rm [OIII]}\sim$2.0$\times$10$^{42}$ \ergs~ (\citealt{Tadhunter2011}) is in the quasar regime (\citealt{Zakamska2003}). Therefore, 4C12.50 is a radio-loud type 2 quasar (QSO2).

4C12.50 is very rich in dust and molecular gas  with a mass of  cold ($\la$25 K) molecular gas   $\sim$10$^{10}$ M$_{\odot}$ (\citealt{Dasyra2012}), in the range of other ULIRGs (\citealt{Mirabel1989,Evans1999,Solomon2005}). It is  the most molecular gas-rich radio galaxy known in the nearby Universe (e.g. \citealt{Ocana2010,Smolvic2011}). Most of this gas is highly concentrated within few kpc of the western active  nucleus, including a small, $\sim$4 kpc wide disk (\citealt{Fotopoulou2019}).

This primary nucleus  hosts  fast  outflows  (up to $\sim$2000 km s$^{-1}$)  which have been  detected  in emission in the  ionized phase and in absorption in the cold HI   circumnuclear gas  (e.g.\citealt{Holt2003,Holt2011,Spoon2009a}, \citealt{Morganti2005,Rose2018}; see also \citealt{Rupke2005a}).  
 The ionized  and the neutral HI outflows in 4C12.50 are  driven by  the radio plasma.  The compact jet seems to be  fighting its way out and emerging from  the dense cocoon of gas and dust in the western nucleus, clearing it out via these kinematically extreme outflows (Morganti et al.   \citeyear{Morganti2003,Morganti2013}).  

It is of great interest to investigate the potential impact of this feedback mechanism in the molecular phase, since this is the fuel used by galaxies to form stars. The evidence for a cold molecular outflow in 4C12.50 is controversial.
Blueshifted CO(3-2) absorption  (shift relative to the systemic velocity $\Delta V\sim$-950 km s$^{-1}$)   and tentative CO(1-0) absorption ($\Delta V\sim$-1100 km s$^{-1}$) (\citealt{Dasyra2012}) have been  reported at  similar velocity as the neutral outflow. On the other hand,  posterior studies (some making use of the Atacama Large Millimeter Array, ALMA)  could not corroborate this outflow, neither  in  absorption or emission using different CO lines  (\citealt{Dasyra2014,Fotopoulou2019,Lamperti2022}). The detection of outflow signatures in the warm ($\sim$400 K)  molecular gas is also controversial. The high velocity blueshifted mid-infrared (MIR) \hmol emission detected by \cite{Dasyra2011}, could not be confirmed by \cite{Guillard2012}.

The outflow  imprint is yet to be searched for in the hot ($>$1500 K) molecular phase. In fact, this  gas, which emits the strongest lines in the near-infrared (NIR),
  has been barely studied in 4C12.50 (\citealt{Veilleux1997,Rose2018}). Our goal is to fill this gap in the knowledge of this otherwise widely studied system. We investigate 
   the general hot \hmol properties (mass, temperature, reddening, excitation mechanisms) and whether the  outflow has affected it.  Moreover, we present new results on the ionized phase, including the high ionization coronal gas. All the results are discussed in the context of prior studies of 4C12.50. This work is based on  Gran Telescopio Canaria (GTC)  K-band and Very Large Telescope (VLT) J, H, K bands long slit spectroscopy of the primary western active nucleus. It is also based on ALMA 12-m array CO(2–1) and 220 GHz continuum  observations described in \cite{Lamperti2022} (program 2018.1.00699.S; PI: M. Pereira-Santaella). See also \citealt{Pereira2021}).

The paper is organized as follows. We describe the observations and data in  \S2. The spectral fitting  techniques are explained in \S3.  The results of the study of the ionized (including coronal)  and the molecular gas components in 4C12.50 are presented in \S4 and discussed in \S5.  The summary and conclusions are in \S6.

We adopt  $H_{0}$=69.7 \kms Mpc$^{-1}$, $\Omega$=0.7185 and $\Omega_{\rm m}$=0.2877. This gives an arcsec to kpc conversion of 2.20 kpc arcsec$^{-1}$  at z=0.122.

\section{Observations}
\label{sec-obs}

\begin{figure}
\centering
\includegraphics[width=0.45\textwidth]{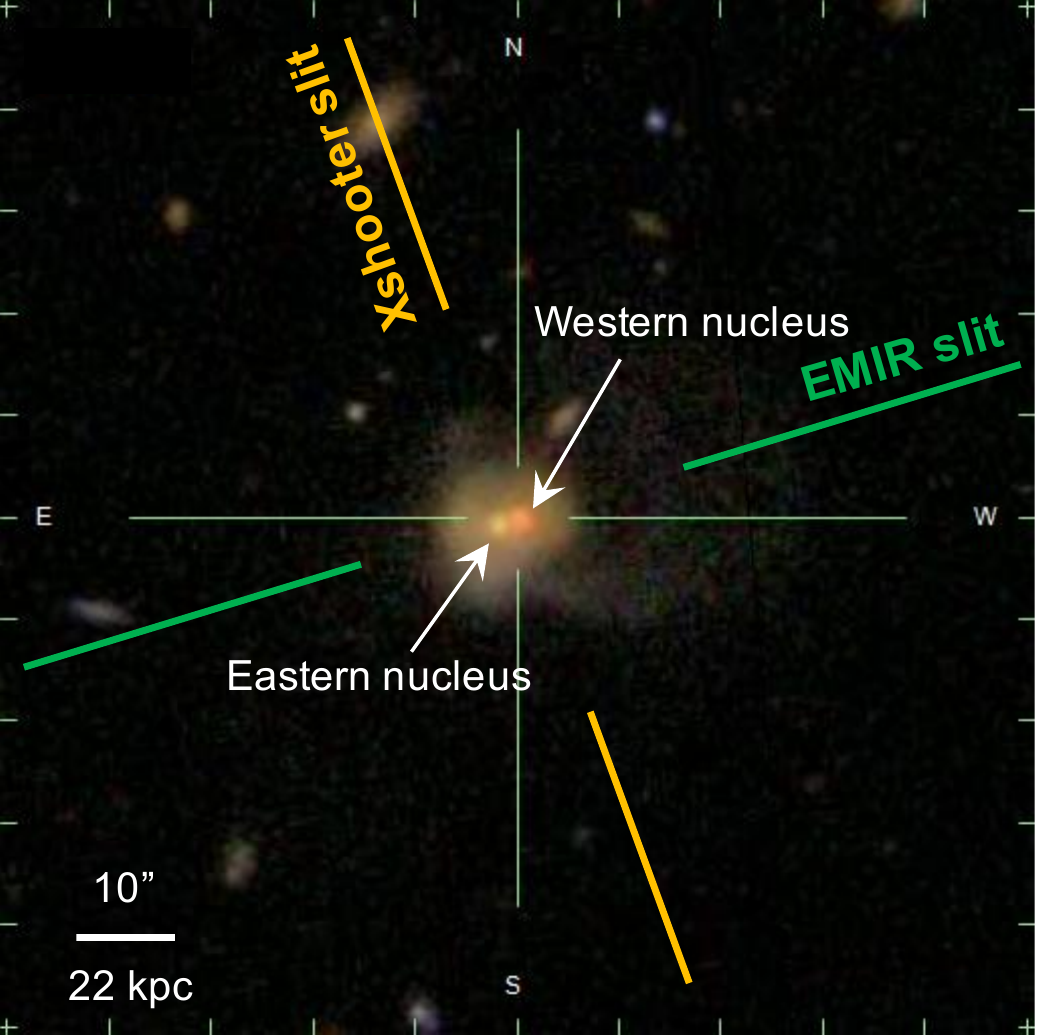}
\caption{SDSS colour composite image of 4C12.50 ($z$=0.122) with the EMIR (position angle, PA=104$\degr$) and Xshooter (PA=20$\degr$) slit positions overplotted. The double nuclei and the prominent merger features are  appreciated clearly. This work is focussed on  the Western active nucleus.}
\label{im1347}
\end{figure}

We obtained K-band spectroscopy of 4C12.50  (RA(2000) 13:47:33.35 and DEC(2000) 12:17:24.2) with the Spanish 10.4m GTC telescope and the EMIR (Espectr\'ografo  Multiobjeto Infra-Rojo)  instrument in long-slit mode (program GTC16-21B). EMIR is a near-infrared 
wide-field imager and medium-resolution multi-object spectrograph installed at the Naysmith-A focal station. It is equipped with a 2048$\times$2048 Teledyne HAWAII-2 HgCdTe   near-infrared optimised  chip with a pixel size of 0.2$\arcsec$. The K  grism covers a spectral range of  $\sim$2.03-2.37 $\mu$m   with a dispersion of 1.71 \AA\  pixel$^{-1}$.

In order to find a compromise between spectral resolution and flux coverage, the slit width used during the observations was 0.8$\arcsec$, adapted to the  K-band  seeing size  (FWHM$\sim$0.8$\arcsec$).  The instrumental profile measured from the arc lines is FWHM$_{\rm IP}$=6.32$\pm$0.44 \AA~ (85.7$\pm$6 km s$^{-1}$ at $\sim$2.2 $\mu$m).  

The slit position angle PA 104$\degr$ N to E was chosen to align the slit with the two  nuclei (Fig. \ref{im1347}). The only line detected from the  secondary, Eastern nucleus is \pa, which is $\sim$90 times fainter than that from the primary nucleus.  For this reason, we will focus our study on the primary nucleus.

Eight spectra were obtained in 4 different nights (Feb 12$^{\rm th}$, April 8$^{\rm th}$, 9$^{\rm th}$,  and 12$^{\rm th}$ of 2022).   The total exposure time on source was 8$\times$3360 sec = 26880 sec or 7.5 hours.  A typical ABBA nodding pattern was applied. 

\begin{figure*}
\centering
\includegraphics[width=0.9\textwidth]{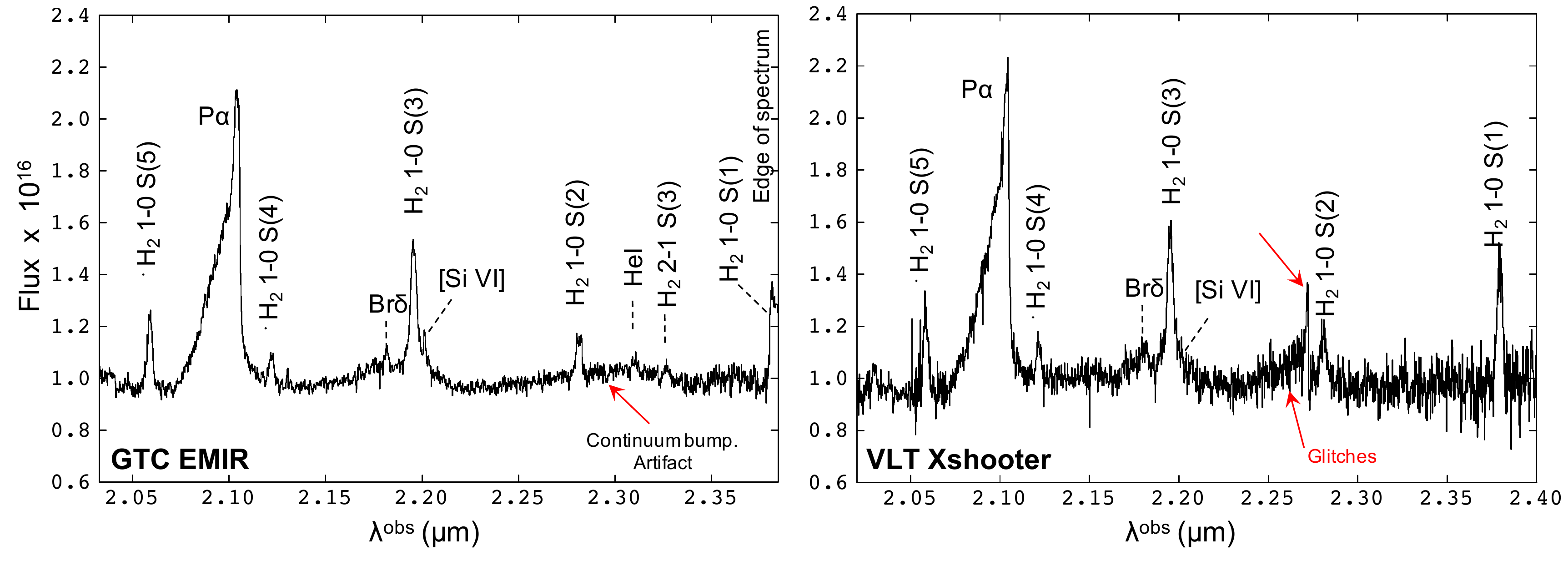}
\caption{K-band spectrum of 4C12.50. Left: GTC-EMIR (0.8$\arcsec\times$1.4$\arcsec$ aperture). Right: VLT-Xshooter (1.2$\arcsec\times$4$\arcsec$ aperture).  Flux  in units of 10$^{-16}$ erg cm$^{-2}$ s$^{-1}$ \AA$^{-1}$. $\lambda^{\rm obs}$ is the observed wavelength. The red arrows mark artifacts and glitches (see text).}
\label{spectra}
\end{figure*}

\begin{figure*}[ !ht ]
\centering
\includegraphics[width=0.8\textwidth]{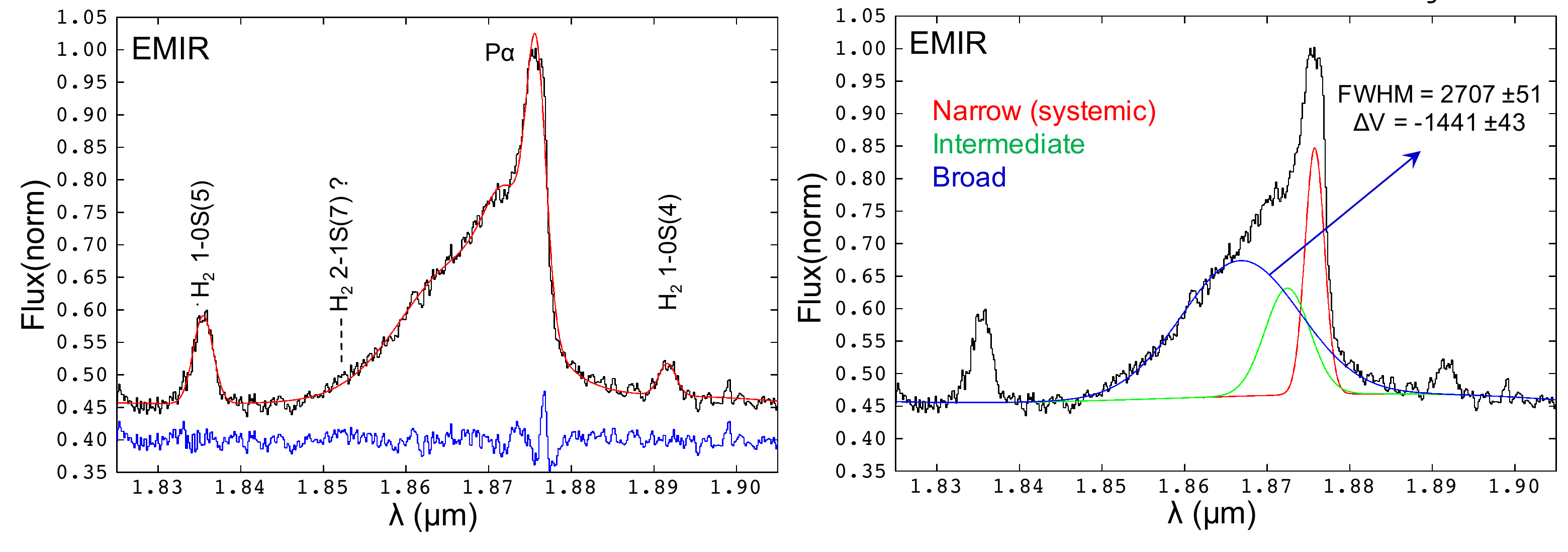}
\includegraphics[width=0.8\textwidth]{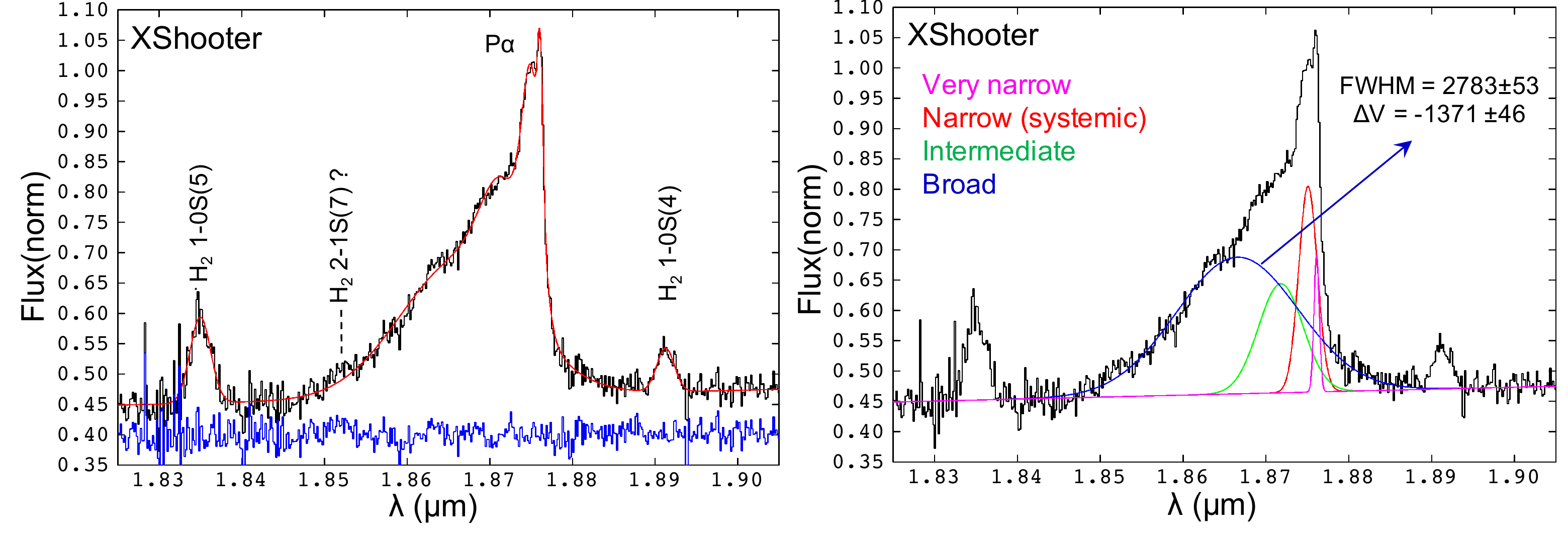}
\caption{Fit of \pa  using the 0.8$\arcsec\times$1.4$\arcsec$ EMIR (top) and 1.2$\arcsec\times$4.0$\arcsec$ Xshooter (bottom) rest frame spectra. Left panels: data (black), fit (red) and residuals (blue), shifted on the vertical axis for visualization. The small excess of flux on the blue wing of \pa  may be H$_2$ 2-1S(7) $\lambda$1.8523. Right:  Individual components of the fit of Pa$\alpha$, where the values of FWHM and $\Delta V$   in \kms~ (velocity shift relative to the narrow or systemic component, in red) refer to the broadest (blue) kinematic component.  It is due to a prominent ionized outflow. The spectra in these and all other figures have been normalized by the peak flux of Pa$\alpha$.}
\label{pa1347}
\end{figure*}

The spectra were reduced using several Python routines customised by GTC staff for EMIR spectroscopic data. The sky background was first subtracted   using consecutive A-B pairs. They were subsequently  flat-fielded, calibrated in wavelength, and combined to obtain the final spectrum. To correct for telluric absorption, we observed a telluric standard star (HR 5238) with the same observing set-up as the science target, immediately  after the 4C12.50 observations and at similar airmass. To apply the correction we used a version of Xtellcor (\citealt{Vacca2003}) specifically modified to account  for the atmospheric conditions of    Roque de los Muchachos observatory in La Palma (\citealt{Ramos2009}). 
Flux calibration was applied using the spectrum of the standard star  obtained with a wide 5$\arcsec$ slit.

The K-band EMIR spectrum is shown in Fig. \ref{spectra} (left).   A 0.8$\arcsec\times$1.4$\arcsec$ aperture centered  at the  continuum centroid of  the western nucleus was chosen. This  optimizes the extraction of the maximum line fluxes and the signal to noise for several important faint lines so that the kinematic parameters can be constrained more accurately.  
The strange continuum bump marked with a red arrow in Fig. \ref{spectra} is an artifact. It suggests a problem with the relative flux calibration in that region of the spectrum. At bluer $\lambda$ the continuum shape is very similar in the eight spectra and the accuracy of the absolute flux calibration  is estimated $\sim$10\%. The shape varies more  towards the red.  The comparison of the line fluxes in the individual spectra suggests an additional $\sim$15$\%$ uncertainty on  the flux calibration in that spectral window (\hmol 1-0 S(2) and redder). This will not affect our conclusions. Moreover, we will also  have the Xshooter NIR spectrum for comparison.

The Xshooter spectrum was  described and shown in \cite{Rose2018}. It  covers  the J+H+K spectral range so that very valuable  information can be obtained from additional emission lines (the K band range is shown in Fig. \ref{spectra}, right).  The spectral resolution values for the 1.2$\arcsec$ slit  were 69.2$\pm$1.3,   70.1$\pm$0.9  and  72.3$\pm$1.3 km s$^{-1}$ for the J, H and K bands respectively. 
 The slit was placed at PA 20$\degr$, the paralactic angle during the observations   (Fig. \ref{im1347}). The pixel scale for the NIR arm is 0.2$\arcsec$. The spectrum was extracted from a 1.2$\arcsec\times$4.0$\arcsec$ aperture, also centered at the western nucleus continuum centroid. The seeing was in the range 0.88$\arcsec$-0.95$\arcsec$ ($g$ band). The authors reported a   $\la$8\% relative flux calibration accuracy.

The Xshooter data has the advantage of covering a much wider spectral range, including  very valuable molecular lines in the J band, as well as 1-0 S(1) (which lies on the edge of the EMIR spectrum). The advantage of the EMIR spectrum is its higher S/N. The r.m.s values in different continuum windows are in the range $\sim$(1.3-2.6)$\times$10$^{-18}$ erg s$^{-1}$  cm$^{-2}$  \AA$^{-1}$, $\sim$1.7-3.5 times lower than for the Xshooter data in the same regions. This spectrum is also more severely affected in some spectral windows by artifacts, including a glitch  in the 2.272-2.283 $\mu$m range (see \cite{Rose2018} for a detailed explanation). Both spectra are therefore valuable.

Galactic extinction is very low ($A_{\rm K}$=0.01) and no correction  was applied for this effect.

In spite of the different aperture sizes, we measure  almost identical  \pa fluxes in the EMIR  ($F_{\rm Pa\alpha}$=(1.6$\pm$0.2)$\times10^{-14}$   erg s$^{-1}$ cm$^{-2}$) and the Xshooter spectra (1.5$\pm$0.1)$\times10^{-14}$ erg s$^{-1}$ cm$^{-2}$).  The same can be said about most (if not all) K-band emission lines (see next section). This suggests that the \pa and other line emission is very compact and strongly concentrated in the primary nucleus and there is little contamination by more extended gas in both apertures.   

\begin{figure}
\centering
\includegraphics[width=0.48\textwidth]{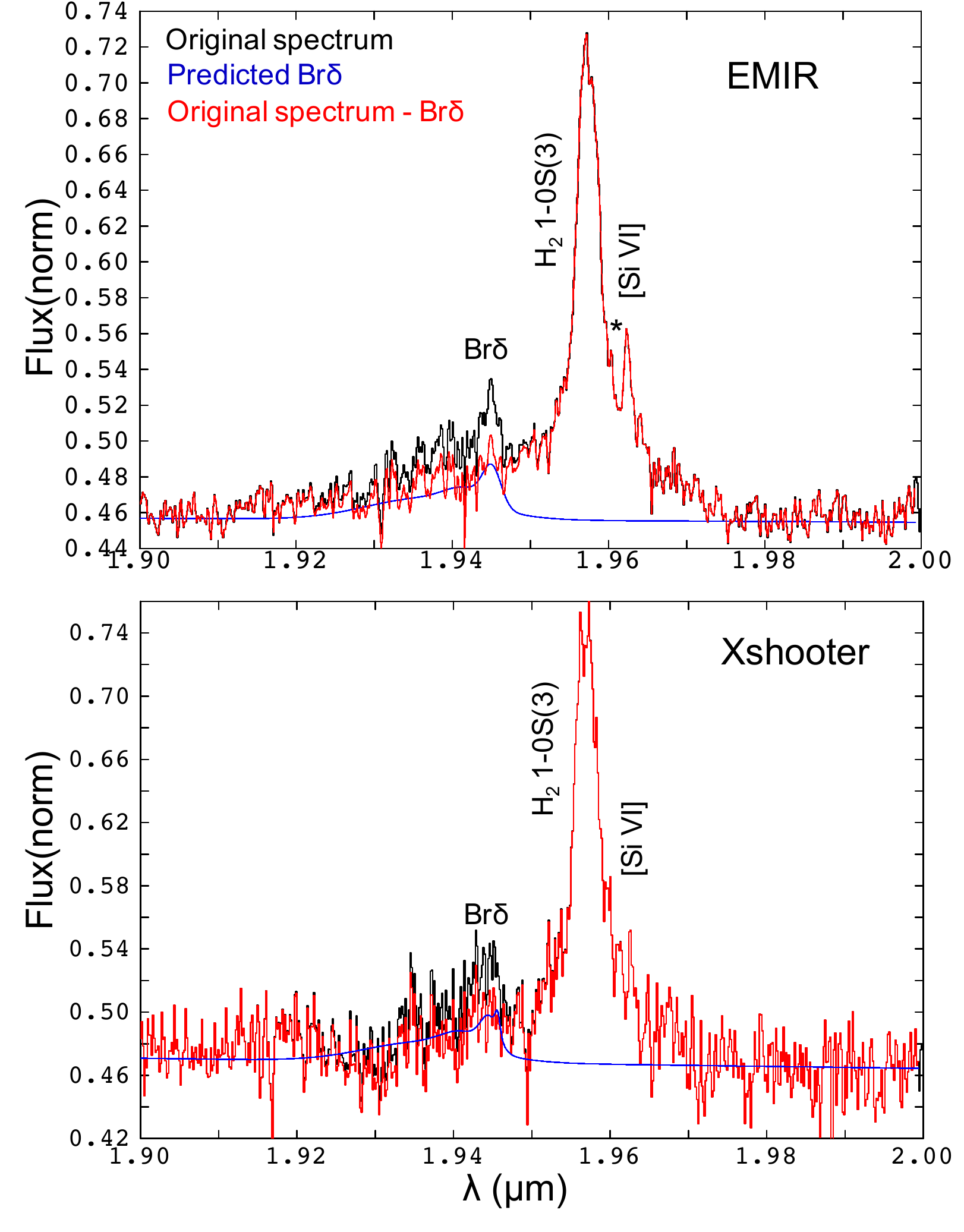}
\caption{Br$\delta$, \hmol (1-0) S(3) and [SiVI] blend (black) in the EMIR (top) and Xshooter (bottom) rest frame spectra.  The blue line shows the predicted Br$\delta$ profile (+ continuum)  (see text). The results of subtracting   Br$\delta$ from the original spectra are shown in red. The * marks the location of  (or near) a possible artifact in the EMIR spectrum, responsible for the unexpectedly narrow [SiVI] peak.}
\label{sivi1}
\end{figure}

\section{Analysis}
\label{analysis}

\begin{table*}[ !ht ]
\centering
\tiny
\begin{tabular}{lcccccccc}
\hline
  &  & \multicolumn{3}{c}{EMIR}&      \multicolumn{4}{c}{Xshooter}\\
\cline{3-5}
\cline{7-9}
\\
Line & $\lambda$ & Flux $\times$10$^{-16}$ & FWHM  & $\Delta V$ &  & Flux $\times$10$^{-16}$ & FWHM & $\Delta V$ \\
& ($\mu$m) &  ergs~cm$^{-2}$~s$^{-1}$ & km~s$^{-1}$       &   km~s$^{-1}$   & & ergs~cm$^{-2}$~s$^{-1}$ & km~s$^{-1}$ &   km~s$^{-1}$     \\ \hline
~Pa$\alpha$(vn)	&	   1.8756 	&    &   &   & &   4.3$\pm$0.6  &  101$\pm$12 &   +167$\pm$9 \\
~~~(n)	&	1.8756 	&  29.5$\pm$1.1  & 438$\pm$9   &  0 &  & 22.8$\pm$2.9 & 419$\pm$18 & 0\\
~~~(i)  &  &   29.4$\pm$2.4 & 1023$\pm$45  &  -532$\pm$24 & & 28.9$\pm$2.8 & 1029$\pm$62 & -528$\pm$35\\
~~~(b) & 	 & 98.2$\pm$3.5  &  2707$\pm$51  & -1441$\pm$43 & &99.0$\pm$4.5  & 2783$\pm$53  & -1371$\pm$46\\
~Pa$\beta$	   &      1.2822	&	out &       out    &    out   &   &52.7$\pm$4.7	&	ns	&	ns\\
~Br$\gamma$	   &      2.1661	&	out &       out    &    out   & &  12.5$\pm$2.5	&	ns	&	ns\\ \hline
\end{tabular}	
\caption{Measurements of H$^+$ lines in the EMIR and Xshooter spectra of 4C12.50.    For Pa$\alpha$, (vn),  (n), (i) and (b) refer to the ``very narrow'' (only confirmed in the Xshooter spectrum), ``narrow'',  ``intermediate'' and ``broad'' kinematic components isolated in the spectral fits.   $\Delta V$ is the velocity shift of the individual kinematic components of \pa relative to the narrow one (n), which is at the systemic $z_{\rm sys}$.   "ns" means  noisy and      "out" means that the line is outside the spectral range.} 
\label{tablelines2}
\end{table*}

\begin{table*}[ !ht ]
\centering
\tiny
\begin{tabular}{lcccccc}
\hline
  &  & \multicolumn{2}{c}{~~~~EMIR} &      \multicolumn{3}{c}{~~~Xshooter}\\
\cline{3-4}
\cline{6-7}
\\
Line & $\lambda$ & Flux $\times$10$^{-16}$ & FWHM  & &   Flux $\times$10$^{-16}$ & FWHM  \\
& ($\mu$m) &  ergs~cm$^{-2}$~s$^{-1}$ & km~s$^{-1}$       &  & ergs~cm$^{-2}$~s$^{-1}$ & km~s$^{-1}$      \\ \hline
~H$_2$ 2-0 S(1)		&	1.1622  & out & out &   & 2.20$\pm$0.21 & 455$\pm$35    \\
~H$_2$ 1-0 S(7)		&	1.7480	 &  out  & out &   & 5.10$\pm$0.37 & 461$\pm$23    \\
~H$_2$ 1-0 S(5)		&	1.8358	 &   11.4$\pm$0.40 &   471$\pm$8  &  & 12.50$\pm$0.60 & 465$\pm$15   \\
~H$_2$ 1-0 S(4) &	1.8920   & 3.65$\pm$0.18   & 424$\pm$17 &  & 4.31$\pm$0.15 & 389$\pm$14  \\
~H$_2$ 1-0 S(3)		&	1.9576	 &      20.1$\pm$0.3   &   418$\pm$12     &  & 16.7$\pm$0.4 &  424$\pm$21 \\
~H$_2$ 1-0 S(2)   &	2.0338  &	 6.03$\pm$0.24 & 453$\pm$9  & & 5.51$\pm$0.84  &   457$\pm$66 (ns)   \\
~H$_2$ 1-0 S(1) &	 2.1218  &       20.0$\pm$1.9    & 446$\pm$26   &  &  21.30$\pm$1.10 &  435$\pm$9  \\ 
 \hline
\end{tabular}	
\caption{Measurements of the \hmol  emission lines detected in the EMIR and/or Xshooter spectra of 4C12.50.   H$_2$ 1-0S(1) is on the red edge of the EMIR spectrum. The S(3) measurements  are based on a simple, 1-Gaussian fit of the line profile above the broad pedestal.} 
\label{tablelines}
\end{table*}

\begin{figure*}
\centering
\includegraphics[width=0.75\textwidth]{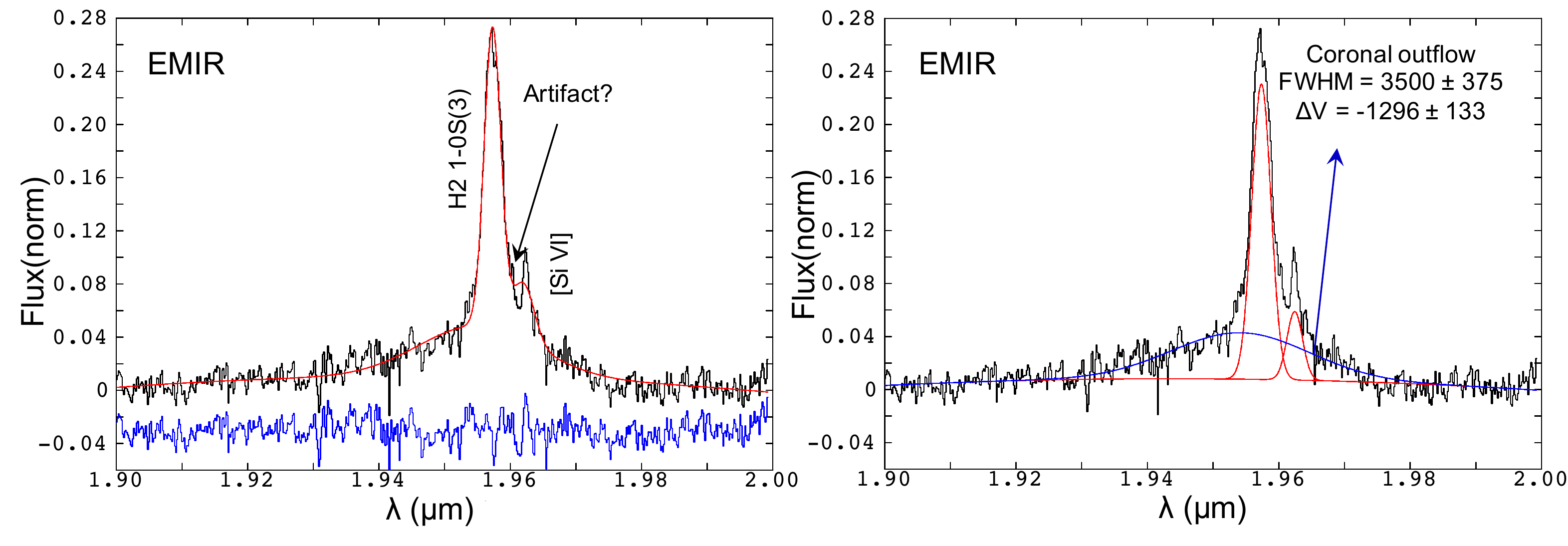}
\includegraphics[width=0.75\textwidth]{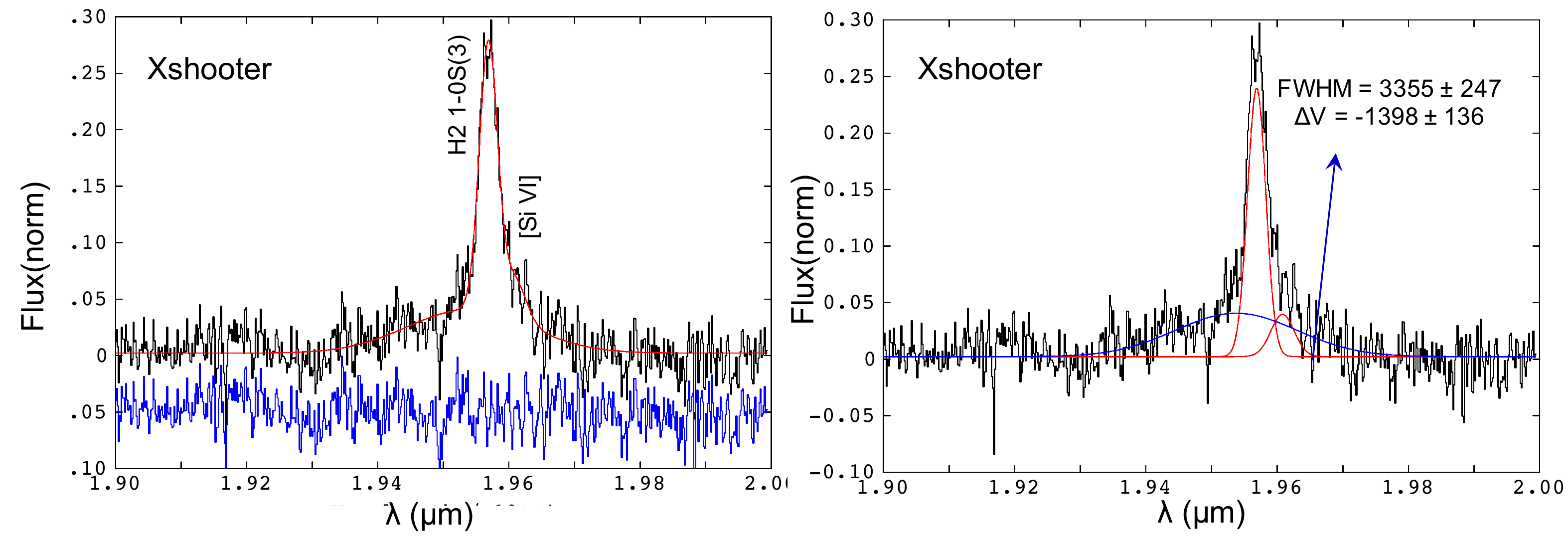}
\caption{Left: Fit of the  \hmol (1-0) S(3) and [SiVI] blend  using the EMIR (top) and Xshooter (bottom) spectra after subracting the continuum and the contribution of Br$\delta$. Color code as in Fig. \ref{pa1347}.  Right: Individual components of the fit. The  FWHM and $\Delta V$  relative to $z_{\rm sys}$ correspond to the broadest component (blue), which is responsible for the broad pedestal}.
\label{sivi2}
\centering
\includegraphics[width=1\textwidth]{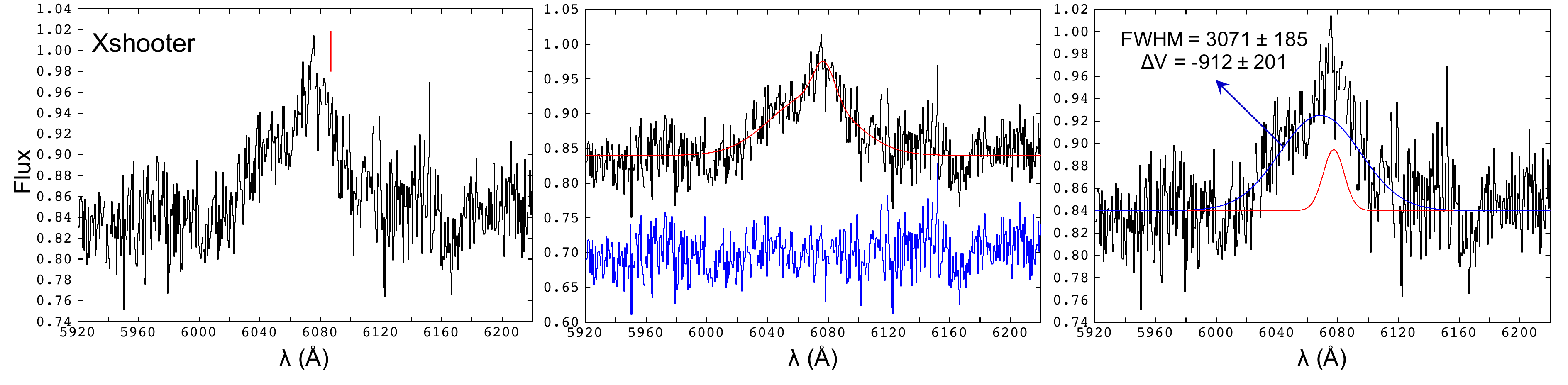}
\caption{Xshooter rest-frame optical spectrum centered on [FeVII]$\lambda$6087. Left: data. The red vertical line marks the expected location of the line for $z_{sys}$. The line is clearly blueshifted. Middle: data with fit (red) and residuals (blue, shifted vertically for visualization). Right: data with the two kinematic components isolated in the fits. The FWHM and $\Delta V$ relative to $z_{sys}$ correspond to the broad most blueshited component. Flux in arbitraty units.}
\label{figfevii}
\end{figure*}

Relevant  H and \hmol  lines detected in the NIR spectrum of 4C12.50 are listed in  Tables \ref{tablelines2} and \ref{tablelines}.  
Except for \pa and the blend of Br$\delta$, \hmol(1-0) S(3) and [SiVI]$\lambda$1.9630 (see below), each line flux  was measured by   integrating the area underneath its spectral  profile and above the continuum level. The FWHM  were obtained from  single Gaussian fits. The {\sc splot} task in {\sc iraf} was used. A single Gaussian  provides acceptable fits in general, even when  slight asymmetries are hinted. 

For complex line profiles  and blends  (see below),
 we applied  multiple-Gaussian fits using the {\sc Starlink} package {\sc dipso}. We used the minimum number of components required to produce an adequate fit, without leaving significant features in the residuals.
{\sc dipso} 
is based on the optimization of fit coefficients, in the sense of minimizing the sum of the squares
of the deviations of the fit from the spectrum data.  The output from a complete fit consists of the optimized parameters (FWHM, central $\lambda$, peak and integrated fluxes) and their
errors (calculated in the linear approximation, from the error matrix). 

All FWHM values  were corrected for instrumental broadening
by subtracting the instrumental profile in quadrature.

\subsection{\pa}
\label{secpa}

\pa  shows a very prominent blue broad wing  (Fig. \ref{pa1347}). 
Three kinematic components are identified in the multi-Gaussian fit of the EMIR spectrum   with FWHM and velocity shifts relative to the narrow component shown in Table \ref{tablelines2}. The results obtained  with the Xshooter spectrum are in good agreement (Table \ref{tablelines2}). The only difference  is that an additional very narrow  component (FWHM=101$\pm$12 km s$^{-1}$  and redshifted by +167$\pm$9 km s$^{-1}$) is isolated in these data set. This can be due to  the higher spectral resolution or the different  position angle and width of the Xshooter slit picking up an additional narrow component.  It only contributes $\sim$2\% of the total line flux. 

 Holt et al. (\citeyear{Holt2003}, see also \citealt{Holt2011,Rodriguez2013,Rose2018}) found  relatively similar complex kinematics for the strongest  optical emission lines from the western nucleus, including forbidden transitions such as [NeV]$\lambda$3426, [OIII]$\lambda\lambda$4959,5007,  [OI]$\lambda$6300 and [SII]$\lambda$6716,6731 ([NeV], [OIII], [OI] and [SII] hereafter). The three components isolated in the [OIII] doublet have  FWHM=340$\pm$23 (narrow), 1255$\pm$12 (intermediate) and 1944$\pm$65 \kms (broad),  with these last two shifted by -402$\pm$9 and -1980$\pm$36 \kms  relative to the narrow, systemic component. \cite{Rose2018} identified an additional, even more extreme component in  the [OIII] lines using the Xshooter spectrum mentioned here. It has FWHM=3100$\pm$200 \kms and $\Delta V$=-331$\pm$60 km s$^{-1}$. This  4th component is not apparent in the \pa fit of either the EMIR or the Xshooter spectra and thus  we will not consider this option.

 The \pa narrow component  has $z$=0.12175$\pm$0.00008. This is consistent with the $z$ of the narrow [OIII]$\lambda$5007 ($z_{\rm sys}$=0.12174$\pm$0.00002). According to \cite{Holt2003}, this component  is at  the  systemic $z$ implied by the stars. It is also consistent within the errors with the CO(2-1) redshift $z$= 0.121680$\pm$0.00004 (\citealt{Lamperti2022}). In what follows, we will also assume that the narrow \pa component is at  $z_{\rm sys}$.

\subsection{The Br$\delta$+\hmol S(3)+[SiVI] blend}
\label{blend}

 A  broad  pedestal is apparent underneath  Br$\delta$ (1.9451$\mu$m), \hmol  (1-0) S(3) and [SiVI]$\lambda$1.9630  (Fig. \ref{sivi1}). Our aim is to discern its nature, since it reveals very high velocity gas that may trace a coronal and/or molecular outflow.
It is also necessary to constrain the flux of the S(3) line.

Due to the complex blend of lines,    it is not possible to apply a multiple Gaussian fit avoiding degeneracies. We have applied a different method consisting of creating an artificial Br$\delta$ that we   subtracted  from the original data, as explained below. We then fitted the residual spectrum, which should retain the contribution of  S(3)  and [SiVI]. This method was applied to both the EMIR and the Xshooter spectra.

To create the expected artificial Br$\delta$ profile for a given spectrum, we assumed the same spectral shape as    Pa$\alpha$. The differential reddening of the three kinematic components (\citealt{Holt2003,Holt2011}) does not affect the line profile significantly. 
For $E_{\rm B-V}=0.59\pm$0.11 (as derived by H$\alpha$/H$\beta$,  Pa$\alpha$/H$\beta$ and  Pa$\beta$/H$\beta$, \citealt{Rose2018}), the expected flux ratio Pa$\alpha$/Br$\delta\sim$18.1 is very close to the case B value, 18.29 (\citealt{Osterbrock2006}).  Thus, we used the fitted \pa profile (see previous section), shifted it  to the \brd~ wavelength and divided its flux by 18.1.  The expected Br$\delta$  contribution  is shown in Fig. \ref{sivi1}, together with the continuum  fitted by interpolating between the blue and red sides  of the line blend.
The results of subtracting Br$\delta$ from the original spectra are. A  broad pedestal is still obvious in both cases.

\begin{table*}
\centering
\begin{tabular}{lcccccccc}
\hline
Line	& 	  $\lambda$  	& Flux	&	FWHM   & $\Delta V$   \\  
	&	$\mu$m &	$\times$10$^{-16}$  \ergcms & \kms	& \kms \\ \hline	
 EMIR & &  \\ \hline
~\hmol S(3) &  1.8756&   21.1$\pm$0.3   &  487$\pm$35    & -42$\pm$9      \\ 
~[SiVI] &    1.9630 &  5.07$\pm$0.88 &  510$\pm$122   &  -120$\pm$44  \\
~Pedestal		 & 1.9630  & 	24.2$\pm$2.2     & 3500$\pm$375  & -1296$\pm$133 \\ \hline
 Xshooter &  & \\ \hline
~\hmol S(3) &  1.8756 &   19.8$\pm$1.8   &  498$\pm$43 &  -90$\pm$21  \\ 
~[SiVI] &  1.9630  & 3.96$\pm$1.64 &  637$\pm$203    & -313$\pm$133  \\
~Pedestal		& 1.9630   &	21.2$\pm$1.8     & 3355$\pm$247  & -1398$\pm$136  \\  \hline
[FeVII] (n) & 0.6087   &  1.28$\pm$0.28 &		 828$\pm$121  & -366$\pm$143\\
~~~(br)   &    0.6087      &   7.06$\pm$0.51 &    3071$\pm$185   &  -912$\pm$201 \\ 
\hline
\end{tabular}	
\caption{Fit of the \hmol (1-0) S(3)+[SiVI] blend using the EMIR spectrum, after removal of the Br$\delta$ contribution (Fig. \ref{fith2sivi}, right) and assuming  the pedestal is dominated by [SiVI] emission. The results of the [FeVII]  spectral fit are also shown for comparison. $\Delta V$ is computed for this line relative to the narrow component of H$\alpha$ as measured in the optical Xshooter spectrum where [FeVII] is detected. }
\label{fith2sivi}
\end{table*}

We then fitted this new spectrum with the smallest number of components (three) that provides a reasonable fit to the residual blend.  The results are shown in Table \ref{fith2sivi} and Fig. \ref{sivi2}. The two narrower components correspond to \hmol S(3) and the core of [SiVI]. 

The  pedestal  is consistent with a very broad  component with FWHM=3500$\pm$375 km s$^{-1}$ for the EMIR spectrum (values for Xshooter will be given in brackets: 3355$\pm$247 km s$^{-1}$). 
 One possibility is that it is  dominated by a molecular outflow, blueshifted by -425$\pm$83 (-478$\pm$136) \kms  relative to the narrow S(3) component. The fact that it is not detected in other molecular lines is not in contradiction.  If its  contribution relative to the narrow component was similar in all \hmol lines ($F_{broad}$/$F_{narrow}\sim$1.15 (1.07), as in S(3)), the expected fluxes would be below or just close to the 3$\sigma$ detection limits in all cases.

While we cannot unambiguously rule out a broad  \hmol S(3) component, the  very turbulent kinematics of the ionized gas (Pa$\alpha$, [OIII], etc)  suggests  that  the   broad pedestal may be dominated  by a coronal [SiVI] outflow.  An intermediate situation, with contribution from both broad S(3) and  [SiVI] or even contamination by  [Si XI]$\lambda$1.9359 on the blue side  of the blend could also be possible.  Overall, this just reflects the difficulty to deblend the \brd +\hmol S(3)+[SiVI] lines.

In spite of this, several arguments support that the pedestal is due to a coronal outflow. In this case,  the line would consist of two components, both blueshifted relative to $z_{\rm sys}$ (Table \ref{fith2sivi}).   The blueshift of both components may indicate that the whole [SiVI] emitting gas (and not only the broad component) is  outflowing, although  asymmetries in the spatial and/or velocity   distributions of the systemic coronal gas cannot be discarded.  The broad component (the pedestal) has   a blueshift of  $\Delta V_{\rm sys}$=-1296$\pm$133 \kms (EMIR) or -1398$\pm$136 km s$^{-1}$ (Xshooter)  and it contributes 83$\pm$11\% of the total [SiVI] flux. 
 This interpretation is supported by the similar kinematics of the coronal [FeVII]$\lambda$6087 line (Fig. \ref{figfevii},  Table \ref{fith2sivi}). The profile is  very broad and asymmetric and clearly blueshifted relative to $z_{\rm sys}$. Two components are isolated, both  broad and blueshifted.  The broadest has FWHM=3071$\pm$185 \kms  and $\Delta V$=-912$\pm$201 \kms. It contributes 85$\pm$9\% of the total line flux.  Thus, [FeVII]$\lambda$6087, as [SiVI], is  dominated by  outflowing gas of rather extreme kinematics.  Finally, \citealt{Spoon2009b} found also very turbulent kinematics for  the MIR coronal [NeV]$\lambda$14.32 $\mu$m (FWHM=2300$\pm$190 and  $\Delta V$=-1120$\pm$89 km s$^{-1}$). Higher S/N and higher spectral resolution would probably reveal a  kinematic substructure  consistent with [FeVII] and [SiVI].

In this scenario,   the ratio of the total line fluxes is \pa/[SiVI]$\sim$5.4. Similar values have been observed in other type 2 quasars (QSO2) and Seyfert 2 (e.g.  \citealt{Riffel2006,Ramos2009,Ramos2019}).  This ratio would be anomalously high ($\sim$31) if the pedestal had no [SiVI] contribution.   The luminosity of [SiVI], $L_{\rm [SiVI]}$=(1.15$\pm$0.07)$\times 10^{41}$ erg s$^{-1}$,  is amongst the highest measured in active galaxies (\citealt{Ardila2011,Lamperti2017,Riffel2006,Cerqueira2021,DenBrok2022}). This is not surprising since most published [SiVI] measurements correspond to  less luminous AGN (Seyfert galaxies), while 4C12.50 is a QSO2 (Sect. \ref{intro}).   SDSS J0945+1737, another QSO2 at $z=$0.128, has a similarly high $L_{\rm [SiVI]}\sim$1.3$\times 10^{41}$ erg s$^{-1}$  (\citealt{Speranza2022}).   \cite{Lamperti2017} found a weak correlation between the [OIII] and [SiVI] fluxes (in log)  for a sample of nearby ($z<$0.075) AGN (see  Fig. 8 in that paper). 4C12.50  is well within the scatter of this relation.

We therefore propose that  the \hmol S(3) emission is traced by the narrow S(3)   above the pedestal, while this (possibly the whole [SiVI] flux) is dominated by a coronal outflow.

\section{Results}
\label{sec-results}

\subsection{The warm ionized and coronal gas}
\label{sec-ion}

The extreme Pa$\alpha$ kinematics are  roughly consistent with those seen in  the optical lines, including the forbidden ones (Sect. \ref{secpa}). This implies that the broad \pa (FWHM$\sim$2750 \kms and $\Delta V$=-1400$\pm$43 \kms (see Table \ref{tablelines}) is not emitted by the broad line region (see also \citealt{Rupke2005a}). It is instead emitted by the kinematically extreme ionized compact outflow, whose radial size $\sim$69 pc was measured by \cite{Tadhunter2018} based on HST narrow band emission line images. 
The  broadest component contributes 63$\pm$3\%  of  the total  \pa flux.
Considering the intermediate component  also as part of the outflow as those authors, this value raises to 81$\pm$4\%.

 \cite{Holt2011}  inferred a very high density $n$ for the broad component. Using the  transauroral emission
lines [S II]$\lambda\lambda$4068,4076 and [O II]$\lambda\lambda\lambda\lambda$7318,7319,7330,7331 they obtain  $n=(3.16^{+1.66}_{-1.01})\times 10^5$ cm$^{-3}$,  compared with $n=(2.94^{+0.71}_{-1.03})\times 10^3$, $(1.47^{+0.60}_{-0.47})\times 10^4$  cm$^{-3}$ for the narrow and the intermediate components respectively (see also \citealt{Rose2018}).   With these values, and using the reddening corrected \pa luminosities we calculate the mass of each kinematic component as:

\begin{equation}
M = \frac{L_{\rm H\beta}~m_{\rm p}}{\alpha_{\rm H\beta}^{\rm eff}~h~ \nu_{\rm H\beta}~n} .
\label{eqn:masstot2}
\end{equation}

\noindent where $L_{\rm H\beta}$ is the the H$\beta$ luminosity,  inferred from the reddening corrected $L_{\rm Pa\alpha}$ and assuming case B $\frac{\rm Pa\alpha}{\rm H\beta}$=0.332 (\citealt{Osterbrock2006}), m$_{\rm p}$ is the mass of the proton, $\alpha_{\rm H\beta}^{\rm eff}$=3.03$\times$10$^{-14}$  cm$^{-3}$ s$^{-1}$ is effective Case B recombination coefficient of H$\beta$ for $T=$10,000 K and $n=$10$^4$ cm$^{-3}$ (\citealt{Osterbrock2006}), $h$ is Planck's constant, $\nu_{H\beta}$ is the frequency of H$\beta$, and $n$ is the electron density. The total  mass is $M_{\rm HII}\sim$9.0$\times$10$^5$ M$_{\rm \odot}$, of which $\sim$6\% corresponds to the broadest component and $\sim$28\% to the total outflowing gas (broad+intermediate components).

\begin{figure}
\centering
\includegraphics[width=0.5\textwidth]{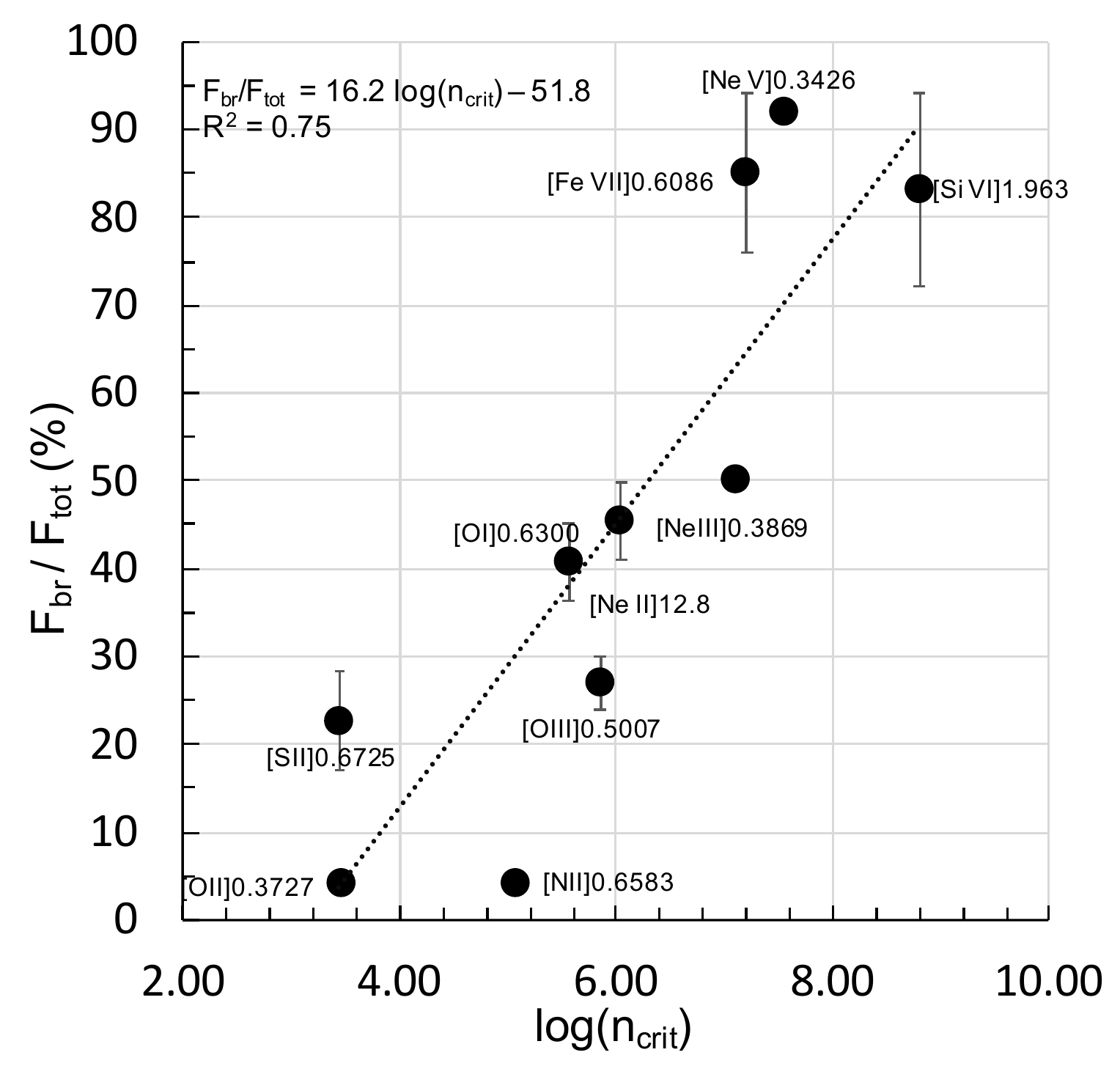}
\caption{Relative contribution  of the broadest component, $\frac{F_{\rm br}}{F_{\rm tot}}$ to the total line fluxes vs. critical density $n_{\rm crit}$ in log and cm$^{-2}$. The dotted line is the linear fit. $\frac{F_{\rm br}}{F_{\rm tot}}$  values are from: this work ([SiVI] and [FeVII]);  \cite{Holt2003} and \cite{Rodriguez2013} (optical lines) and \citealt{Guillard2012} ([NeII]12.8$\mu$m).  Data with no error bars have no errors available.}
\label{fig-ncrit}
\end{figure}

We plot in Fig. \ref{fig-ncrit} the observed relative contribution  of the broadest component to the total line fluxes $\frac{F_{\rm br}}{F_{\rm tot}}$ against the critical density $n_{\rm crit}$ for all the forbidden lines with this information available. Although with a large scatter, a correlation is clear\footnote{The correlation is weaker with ionization potential. It is not shown for simplicity.}. 
The  emission from the most turbulent gas is  much  stronger (dominant) in  lines with high  $n_{\rm crit}$ such as the coronal lines than in  low  $n_{\rm crit}$ lines, specially those with $n_{\rm crit}$  lower than the outflow density (e.g. [OII]$\lambda$3727 with $n_{\rm crit}$=1300 and 4500 cm$^{-3}$ for the two doublet components and [SII]$\lambda\lambda$6716,6731, $n_{\rm crit}$=1500 and 4000 cm$^{-3}$).   This is consistent  with the frequent finding that AGN outflows have a more prominent signature in coronal features compared with  other lines from both the ionized and specially the molecular phases (\citealt{DeRobertis1984,Ardila2002,Alvarez2023}).  

A similar correlation was found in  MRK477, the nearest QSO2 at $z=$0.035  by \cite{Villar2015}.   The NLR density is expected to decrease with distance from the AGN (e.g. \citealt{Bennert2006}; see also \citealt{DeRobertis1984}). The authors proposed that the  outflow in MRK477 has been triggered  at $\la$220  pc, possibly at $\la$30 pc, from the AGN (also by the radio jet) and the correlation shows how its emission weakens as it propagates outwards from the inner denser coronal region (e.g. \citealt{Muller2006,Rose2011}) outwards in the NLR, following the decreasing density gradient. We propose a similar scenario  for 4C12.50. 

 This also suggests that a single mechanism (the radio jet in this case) is responsible for the outflow identified in all emission lines emitted by the ionized phase, from the coronal to the lowest ionization species.  This is also supported by the  similar, unusually high  values of $V_{\rm max} =|\Delta V_{\rm sys} - FWHM/2|$ (a frequent definition of the maximum outflow velocity, computed for the most blueshifted component; \citealt{Rupke2005b}) for most lines (Table \ref{tab-vmax}). 

\begin{table}
\centering
\begin{tabular}{lcc}
\hline
Line	&		 $V_{\rm max}$\\ 
	&	 \kms \\ \hline
~Pa$\alpha$ & 2763$\pm$53$^a$  \\
~[SII]$\lambda\lambda$6716,6731 &  2803$\pm$242$^b$  \\
~[NeII]$\lambda$12.8	& 	1831$\pm$137$^c$  \\
~[OIII]$\lambda$5007 &   2952$\pm$48$^b$ \\
~[OI]$\lambda$6300   &		 2431$\pm$80$^b$  \\
~[FeVII]$\lambda$6087 &		 2496$\pm$221$^a$ \\
~[SiVI]$\lambda$1.9623 & 3076$\pm$184$^a$ &cr\\
\hline
\end{tabular}	
\caption{$V_{\rm max} =|\Delta V_{\rm sys} - FWHM/2|$ (\citealt{Rupke2005b}) of the broadest component for  several emission lines. Values from $^a$this work, $^b$\cite{Holt2003}, $^c$\citealt{Guillard2012}. The lines below \pa are in order of incresing $n_{\rm crit}$.}
\label{tab-vmax}
\end{table}

The  narrow  \pa component is at the systemic redshift  (Sect. \ref{secpa}). 
The prominent  blue  excess and the almost total lack of redshifted emission   (Fig. \ref{pa1347}) could be due to  an asymmetric spatial distribution of the outflowing gas. This is  expected in 4C12.50, since the  interaction between the radio source  and the ambient gas at both sides of the AGN is indeed very asymmetric (\citealt{Morganti2004}).  
  
Another possibility is that   the  receding side of the outflow  is almost completely extinguished by dust.
 Let us consider the broadest, most turbulent component.  If the gas moving away on the far side of the nucleus has the same kinematics and emits the same intrinsic \pa flux as the approaching side ($\sim$2.1$\times$10$^{-14}$ erg s$^{-1}$ cm$^{-2}$, corrected for reddening with $E_{\rm B-V}$=1.44 mag, \citealt{Holt2003}), the comparison with the 3$\sigma$ upper limit $\la$9.4$\times$10$^{-16}$ erg s$^{-1}$ cm$^{-2}$, implies $E_{\rm B-V}>$5.8 or $A_{\rm V}>$23.5 mag.  Assuming the same  gas-to-extinction ratio as in our Galaxy (\citealt{Zhu2017}), this corresponds to  a column density of HI, 
$N_{\rm HI} \sim$2.08$\times$10$^{21} \times A_{\rm V}$  cm$^{-2}\ga$4.9$\times$10$^{22}$ cm$^{-3}$, which is consistent with   $N_{\rm HI}$ measurements in the central region ($\sim$200 pc) of 4C12.50 (\citealt{Morganti2013}).

Therefore, it is possible that the receding ioinized outflow is completely extinguished. This would not be surprising, given  the dusty circumnuclear environment of this and other ULIRGs and the  presence of a circumnuclear torus related to the 
AGN. The  outflow, therefore, could be trully kinematically  extreme in this case, with FWHM$\sim$2$\times$FWHM$_{\rm broad}\sim$5400 \kms ($\sim$7000 \kms for the coronal gas, Table \ref{fith2sivi}), only comparable to those seen in a handful of high $z$ extremely luminous quasars (\citealt{Perrotta2019,Villar2020}). 
The  outflow  mass, $M\sim$2.5$\times$10$^5$ \msun (broad and intermediate \pa components, see above), the mass outflow rate, $\dot{M}$ and  the kinetic power $\dot E_{\rm kin}$ would still be moderate. These  are  calculated as (e.g. \citealt{Rose2018}): 

\begin{equation}
\dot{M} =\frac{M~V}{r} .
\label{eqn:mass}
\end{equation}

\begin{equation}
\dot{E} =\frac{\dot{M}}{2} V^{2} ,
\label{eqn:pow}
\end{equation}

where  $V$ is the average velocity of the outflowing gas and $r$ is the outflow radius. We assume $V$ = $V_{\rm max}$ (\citealt{Rupke2005b}).  For a Gaussian of a given FWHM centered at $V_{\rm sys}$, $V_{\rm max}$=FWHM/2. In the current scenario,   $V_{\rm max}\sim$2783 \kms and$\sim$1029 \kms  for the broadest and intermediate components respectively  (Table \ref{tablelines2}). Assuming $r$=69 pc (\citealt{Tadhunter2018}),  then $\dot{M}\sim$3.3 \msun yr$^{-1}$  and  $\dot{E}\sim$1.1$\times$10$^{43}$ erg s$^{-1}$ for the broad component; $\dot{M}\sim$4.4 \msun yr$^{-1}$  and  $\dot{E}\sim$2.0$\times$10$^{42}$ erg s$^{-1}$ for the intermediate component. In total, $\dot{M}\sim$7.7 \msun yr$^{-1}$  ($\ll$SFR$\sim$100 \msun yr$^{-1}$, \citealt{Rupke2005a}) and  $\dot{E}\sim$1.3$\times$10$^{43}$ \ergs, which is $\sim$0.14\% of the bolometric luminosity $L_{\rm bol}\sim$9$\times$10$^{45}$ ergs. The values are still moderate  (see also \citealt{Holt2011,Rose2018}).  Given, in addition, the small size  (much smaller than the effective radius of the western bulge component, 2.59$\pm$0.58 kpc, \citealt{Dasyra2006}) and volume apparently affected by the ionized outflow,  it is not clear whether  it will affect the evolution of the host galaxy.

\subsection{The hot molecular gas}
\label{sec-mol}

\subsubsection{Rotational temperature, $T_{\rm rot}$}
\label{tempmol}

We have calculated the \hmol rotational excitation temperature, $T_{\rm rot}$, using  the extinction corrected fluxes and upper limits of the   NIR \hmol lines and following  \cite{Pereira2014}.  The molecular lines
 are not necessarily affected by the same extinction as the ionized gas emission.  To estimate $E_{\rm B-V}^{\rm mol}$, we have used $\frac{\rm H_2~ 2-0~ S(3) \lambda 1.1175}{\rm H_2 ~2-1~ S(3)\lambda 2.0735}$, which  has a theoretical  value of 0.83. Both lines are detected in the Xshooter spectrum. Because they are faint and noisy, different aperture sizes were attempted to maximize the S/N.  We infer a value of 0.58$\pm$0.05. This implies $E_{\rm B-V}^{\rm mol}$=0.35$\pm$0.08, which is not significantly different in comparison with  the ionized gas total extinction E(B-V)=0.59$\pm$0.11 (\citealt{Holt2003,Rose2018}).

\begin{figure}
\centering
\includegraphics[width=0.48\textwidth]{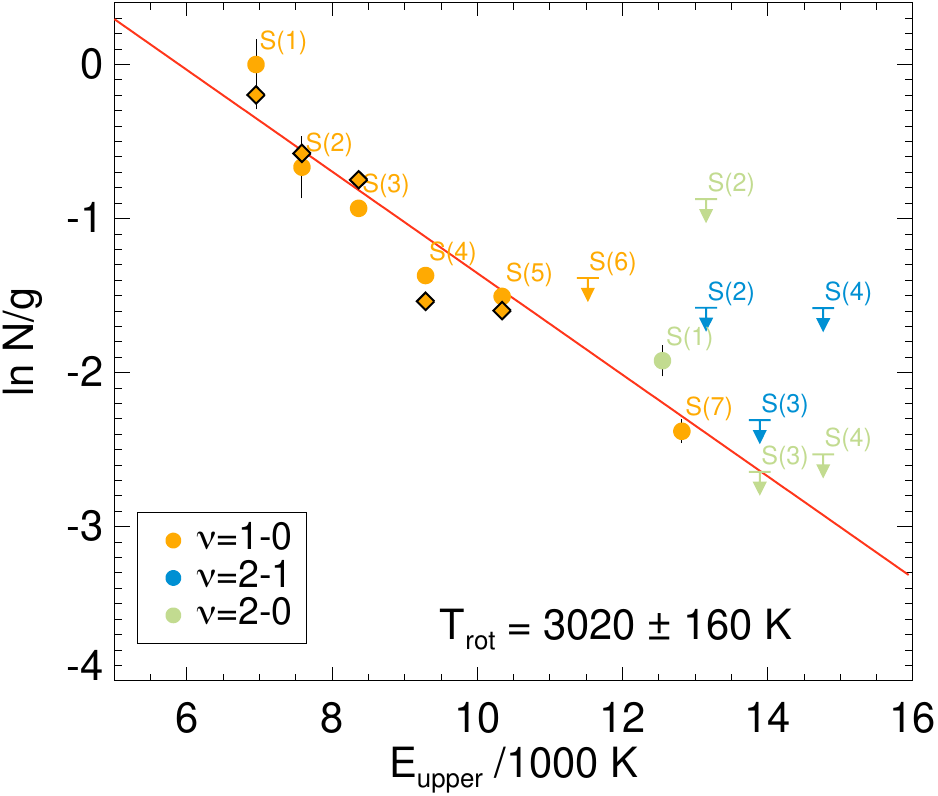}
\caption{Modelling the relative population levels of the \hmol NIR transitions using single excitation-temperature LTE models. The orange circles and diamonds represent de Xshooter and EMIR measurements respectively. Upper limits were obtained with the Xshooter spectrum. They correspond to  \hmol 1-0 S(6) $\lambda$1.7880  (orange), the 2-1 transitions (blue) S(2)  $\lambda$2.1542, S(3) $\lambda$2.0729 and S(4) $\lambda$2.0041  and the \hmol 2-0 transitions (light green) S(2)  $\lambda$1.1382, S(3)  $\lambda$1.1175 and S(4)  $\lambda$1.0998 in the J band. The solid red line shows the single-temperature fit based on the  fluxes of the lines detected in the Xshooter spectrum and assuming fully thermalized LTE gas conditions. The EMIR data are shown for comparison.}
\label{figtemp}
\end{figure}

We show in Fig. \ref{figtemp} the result of modelling the relative population levels of the NIR   \hmol lines using  single excitation-temperature LTE models (see \citealt{Pereira2014}).  
We infer $T_{\rm rot}=$3020$\pm$160 K. This is quite high in comparison with typical values in nearby AGN and ULIRGs. As an example, \cite{Riffel2021} obtained  $T_{\rm rot}$ in the range $\sim$760-2075 K in a sample of  36 nearby Sy1 and Sy2 (0.001$\la z \la$0.056) selected  among the hard X-ray (14–195 keV) sources in the Swift
Burst Alert Telescope (BAT) survey (\citealt{Oh2018}). The ULIRGs analysed by \cite{Davies2003}, which also host very large  $M_{\rm hot}$ (see next section), have  maximum temperatures of $\sim$2400 K. 
 In general, $T_{\rm rot}<$2500 K in galaxies, including U/LIRGs and AGN, (\cite{Murphy2001,Davies2003,Ardila2004,Ardila2005,Ramos2009,Mazzalay2013,Pereira2014}.  Such high temperature is not found either in hot molecular outflows ($T_{\rm rot}\sim$1900-2300  K), although there is only a handful of   objects where it has been possible to isolate the  NIR \hmol outflow emission (\citealt{Emonts2014,Tadhunter2014,Ramos2019}). 

The NIR \hmol lines extend the  gradient found by \cite{Guillard2012} in 4C2.50 to higher temperatures. They fitted three  components using  the MIR \hmol lines  with  $T\sim$100, 275 and 1500 K respectively.   A temperature   gradient, described with a power law, exists in the molecular gas of numerous galaxies (\citealt{Davies2003,Ogle2010,Guillard2012,Pereira2014,Togi2016}).

\subsubsection{Mass}
\label{massmol}

The mass of hot molecular gas $M_{\rm H_2}^{\rm hot}$ can be estimated from the extinction corrected S(1) flux, $F_{\rm H_2 1-0 S(1)}$,   under the assumptions of local thermal equilibrium and an excitation temperature of 3020$\pm$160 K  (e.g. \citealt{Scoville1982,Riffel2014}). We infer $F_{\rm H_2 1-0 S(1)}$=(2.44$\pm$0.41)$\times$10$^{-15}$  erg s$^{-1}$ cm$^{-2}$   (we use the Xshooter line flux in this calculation because it is more accurate), assuming E$_{\rm B-V}$=0.35$\pm$0.08 (see Sect. \ref{tempmol}).  If the gas was not thermalized, the mass would be underestimated. 

\begin{figure}
\centering
\includegraphics[width=0.52\textwidth]{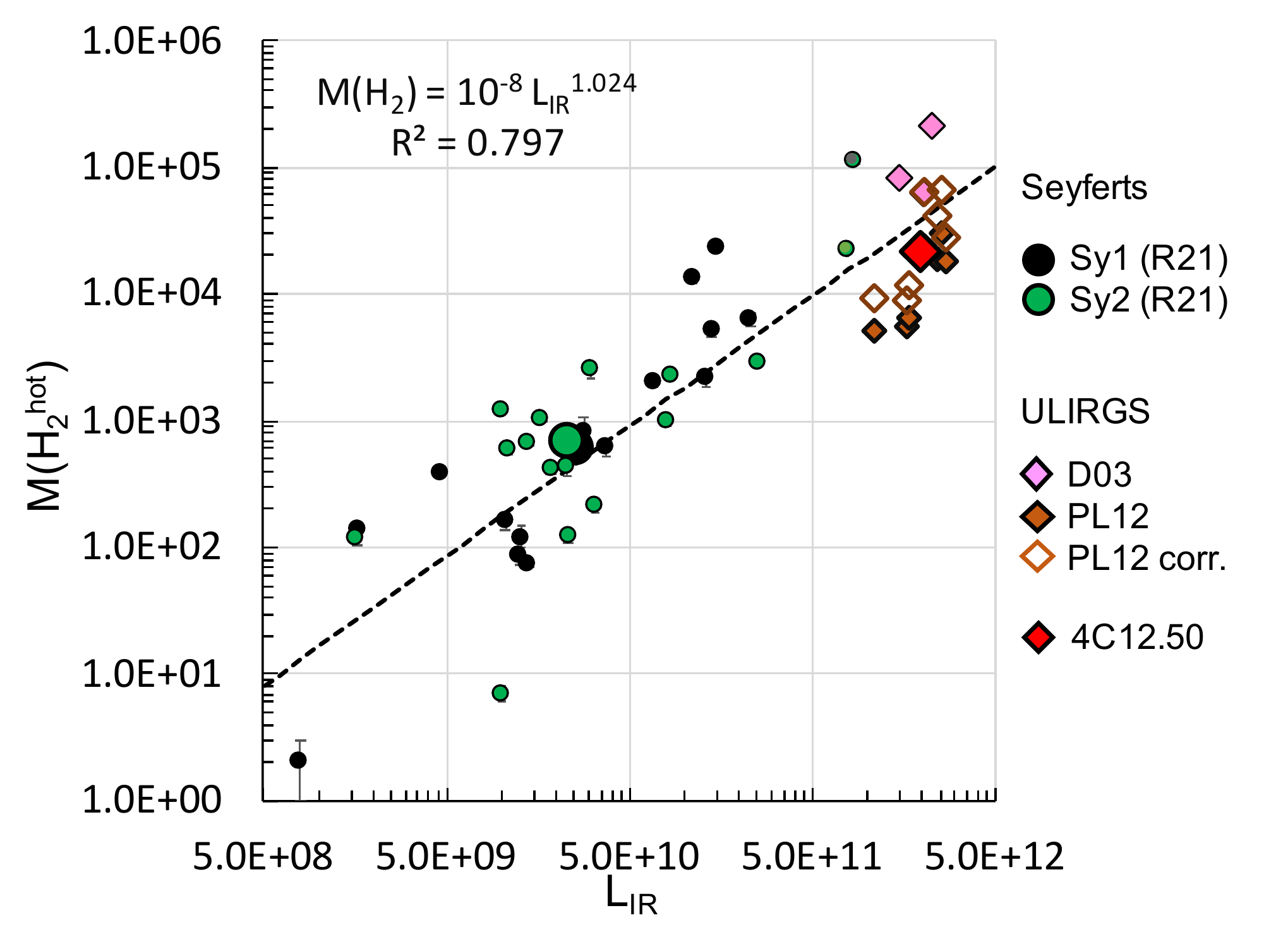}
\caption{$M_{\rm H_2}^{\rm hot}$ in units of $M_{\rm \odot}$ vs.  the 8-1000 $\mu$m infrared luminosity $L_{\rm IR}$  in units of $L_{\rm \odot}$ for  the sample of Sy1 and Sy2 from R21 (\citealt{Riffel2021}) and ULIRGs from D03 ( \citealt{Davies2003}) and PL12 (\citealt{Piqueras2012}). No extinction correction has been applied, except for L12 ULIRGs (open orange diamonds; PL12 corr.).  The same extinction as the ionized gas has been assumed (\citealt{Piqueras2013}). 4C12.50 is plotted as a red diamond. The observed and extinction corrected masses are very similar. The size of the symbol is similar to the errobar. The large black and green solid circles (in an almost identical location) are the median  of the Sy1 (small black circles) and Sy2 (small green circles) values. The dotted black line shows the best fit to all  solid data points (this is, the masses not corrected for extinction).}
\label{mvslir}
\end{figure}

We obtain  $M_{\rm H_2}^{\rm hot}=$  (2.10$\pm$0.44$)\times$10$^{4}$ M$_{\rm \odot}$. 
 We show in Fig. \ref{mvslir} $M_{\rm H_2}^{\rm hot}$  vs. $L_{\rm IR}$ for  \citealt{Riffel2021} sample of nearby Seyfert 1 and 2 previously mentioned (Sect. \ref{tempmol}). Mass values are also shown for several  ULIRGs from \cite{Davies2003} and \cite{Piqueras2012}.    $M_{\rm H_2}^{\rm hot}$ of 4C12.50  is at the high end of values  found in  galaxies, including active galaxies and U/LIRGs (see also \citealt{Ardila2005,Piqueras2012,Mazzalay2013,Mezcua2015,Riffel2021}). It is similar to other nearby ULIRGs and 
consistent with the value expected from  the observed $M_{\rm H_2}^{\rm hot}$ vs. $L_{\rm IR}$ correlation.  

If  slit losses were  significant, the intrinsic  mass would be  higher, although this situation is unlikely.
 The hot molecular gas in luminous AGN is usually mostly concentrated within   $\la$several$\times$100 pc (\citealt{Mezcua2015,Riffel2021}). At the $z$ of 4C12.50, such physical size is not resolved spatially in our data.    Moreover, since the  K-band seeing (see Sect. \ref{sec-obs}),  was significantly narrower than the Xshooter 1.2$\arcsec$ slit,  losses are expected to be low in the direction perpendicular to the slit also.

With log($L_{\rm IR}/L_{\rm \odot}$)=12.31 and $L'_{\rm CO}$=(1.25$\pm$0.38)$\times$10$^{10}$ K km s$^{-1}$ pc$^2$ (\citealt{Dasyra2012}), 4C12.50 lies close to   the $L'_{\rm CO}$ vs. $L_{\rm IR}$ correlation for galaxies and, as other local ULIRGs, it is slightly below it (e.g. Fig. 5 in \citealt{Cortzen2019})). Thus, $M_{\rm H_2}^{\rm cold}$=(1.0$\pm$0.1)$\times$10$^{10}$ \msun (\citealt{Dasyra2014})  is also consistent with that expected for its $L_{\rm IR}$  (see Fig. 1 in \citealt{Daddi2010a}).

\begin{figure}
\centering
\includegraphics[width=0.48\textwidth]{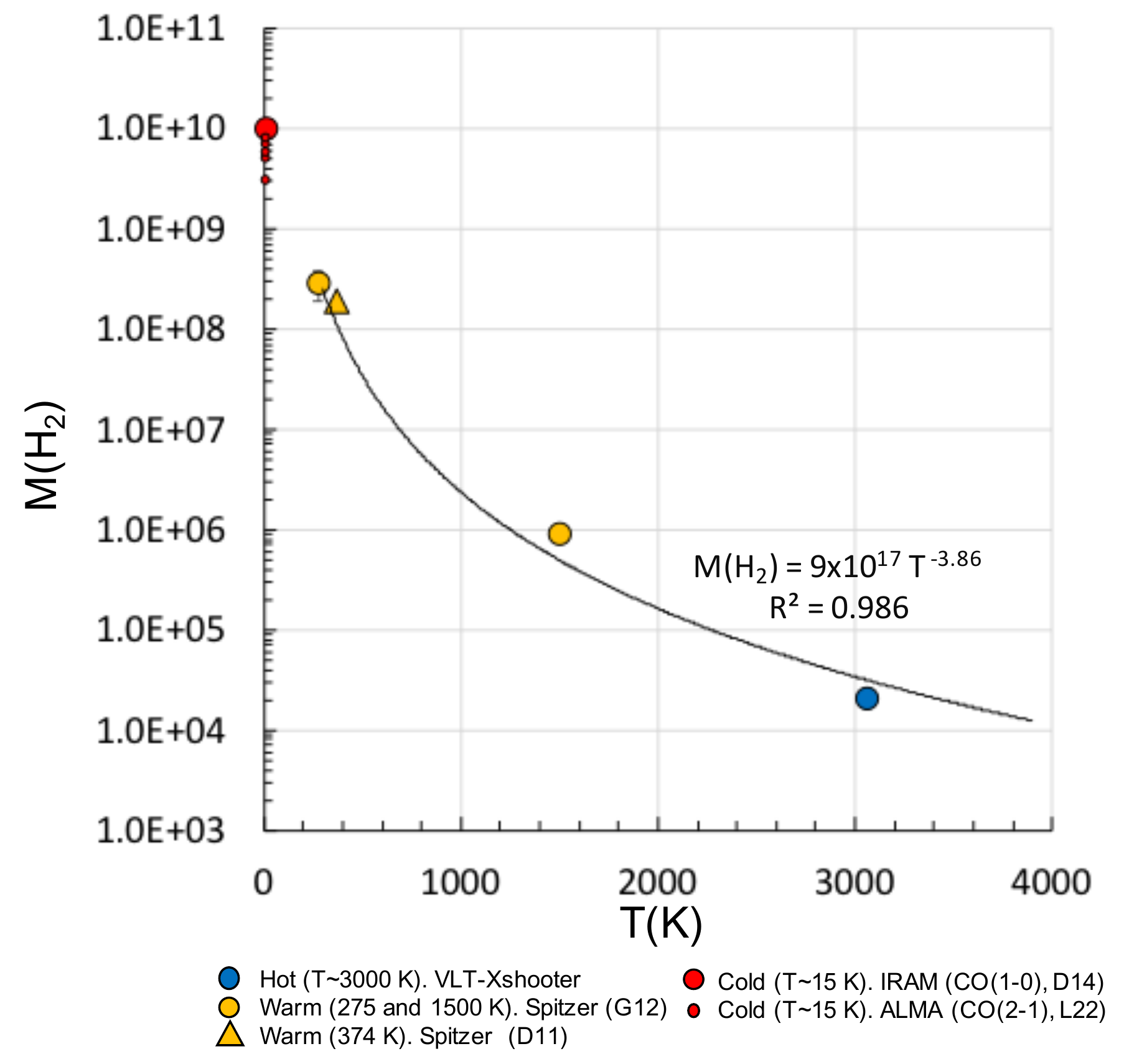}
\caption{Dependence of the molecular gas mass  in \msun with excitation temperature  for 4C12.50. The color code clarifies the tempertature ranges considered as "cold", "warm" and "hot" in this work. The small red circles indicate the cold \hmol masses infered from CO(2-1) ALMA data for different apertures with the radius varying from 0.5$\arcsec$ to 2$\arcsec$ (see text). The black solid  line shows the best fit to the warm (G12) and hot data points, with the equation and coefficient of determination, $R^2$. G12: \cite{Guillard2012}; D11: \cite{Dasyra2011}; D14: \cite{Dasyra2014}; L22: \cite{Lamperti2022}.}
\label{mass-temp}
\end{figure}

The  ratio of  hot  to cold \hmol masses, $\frac{M_{\rm H_2}^{\rm hot}}{M_{\rm H_2}^{\rm cold}}$ = (2.10$\pm$0.49)$\times$10$^{-6}$ M$_{\rm \odot}$, is within the range ($\sim$[10$^{-7}$-10$^{-5}$]) observed across a sample of several dozen star-forming galaxies and AGN by \cite{Dale2005}. It   is moreover consistent with the value expected from its $\frac{f_{60 \mu m}}{f_{100 \mu m}}$=0.93 IRAS color (see Fig. 4 in \citealt{Dale2005}).

The dependence of  $M_{\rm H_2}$ with the excitation temperature, $T$, is shown in   Fig.\ref{mass-temp}. 
The mass at $T\sim$3000 K follows the trend of this and other galaxies, where the  bulk of the  molecular mass  is concentrated at the lowest temperature (\citealt{Guillard2012}). The mass-temperature function at $T\ga$300 K is well described in 4C12.50 with a power law: $M_{\rm H_2}$(\msunb) = 9$\times$10$^{17} ~T^{-3.9}$ (coefficient of determination, $R^2$=0.937) or $\frac{d M_{\rm H2}}{dT}\propto T^{-n}$ with $n\sim$5.
Aperture effects do not have a significant impact in this plot. 
Although the warm masses (\citealt{Guillard2012}) were obtained with  Spitzer IRS data (which provides much larger aperture sizes (3.6$\arcsec \times 57\arcsec$ to 11$\arcsec \times 22\arcsec$, depending on the line), the gas is expected to be highly concentrated within a spatial region smaller than the physical region covered by the  Xshooter aperture (2.6$\times$8.8 kpc$^2$).

 A mass-temperature power law distribution of the molecular gas at T$\ga$100 K  is frequently observed (or assumed) for galaxies independently of the excitation mechanism.   \cite{Togi2016} found that  a continuous power-law distribution of rotational temperatures, with $\frac{d M_{\rm H2}}{dT}\propto T^{-n}$,  reproduces well the \hmol excitation from a wide range of galaxy types using a single  parameter, the power-law slope $n$ (see also \citealt{Pereira2014}).   This model, can recover the mass at $T\ga$100 K,  with  $n$ in the range 3.79-6.4 and average 4.84$\pm$0.61.  $n$ gives information on the relative importance of gas heating by shocks, photoelectric heating, UV pumping, etc. According to \cite{Neufeld2008} $n\sim$4-5  is consistent with the predictions of simple models for paraboloidal bow shocks.

 For 4C12.50, the high $T_{\rm rot}$ and the $n\sim$5 power law mass-temperature distribution suggest that shocks play an important role on the excitation of the molecular gas  at T$\ga$300 K.

\subsubsection{Excitation mechanism}
\label{excitmol}

We have seen that the high $T_{\rm rot}$ and the power law mass-temperature distribution suggest that shocks play an important role on the excitation of the molecular gas at T$\ga$300 K. We now check whether the influence of shocks is apparent in the  diagnostic diagram [Fe II] $\lambda$1.257$\mu$m/Pa$\beta$   vs. \hmol 1-0 S(1)/Br$\gamma$ (\citealt{Larkin1998,Ardila2005,Riffel2021}; see also \citealt{Colina2015}).   [Fe II] $\lambda$1.257$\mu$m/Pa$\beta$=0.73$\pm$0.08 for 4C12.50 is obtained from the Xshooter spectrum.

 Br$\gamma$ is outside the observed spectral range of the EMIR data and very noisy in the Xshooter spectrum. We measure $F_{\rm Br\gamma}$=(1.25$\pm$0.25)$\times$10$^{-15}$  erg 
s$^{-1}$ cm$^{-2}$ (Table \ref{tablelines2}) and thus,   S(1)/Br$\gamma$=1.70$\pm$0.35.  $F_{\rm Br\gamma}$ agrees  within the errors with the reddened flux  predicted from \pa assuming E(B-V)=0.59$\pm$0.11 (\citealt{Holt2003,Rose2018}): $F_{\rm Br\gamma}$=(1.43$\pm$0.08)$\times$10$^{-15}$  erg  s$^{-1}$ cm$^{-2}$ and, thus, S(1)/Br$\gamma$=1.49$\pm$0.11. 
 
4C12.50 is in the  area of the  [Fe II] $\lambda$1.257$\mu$m/Pa$\beta$   vs. \hmol 1-0 S(1)/Br$\gamma$ occupied by AGN.    According to \cite{Riffel2021},  0.4$\le$ \hmol  S(1)/Br$\gamma<$2 for low excitation AGN; 2$\le$  S(1)/Br$\gamma<$6 for high excitation AGN and S(1)/Br$\gamma>$6 for shock dominated regions/objects. Thus, the ratio  is typical of high excitation AGN and below values expected for shock dominated regions.

This, however, does not imply that  shocks are not present and a more likely scenario is that a combination of excitation mechanisms exists.   4C12.50 is part of the MOHEG (molecular hydrogen emission
galaxy) sample of radio galaxies hosting fast ionized and HI jet-driven outflows studied by \cite{Guillard2012}. 
 They  discarded AGN X-ray heating as the dominant source of excitation of the MIR \hmol in these systems,  based on the large ratio of the \hmol line luminosity (summed over the MIR S(0) to S(3) rotational transitions) to the unabsorbed 2-10 keV nuclear X-ray luminosity (see also \citealt{Ogle2010}).  Instead, shocks are proposed as the main    excitation mechanism. Using  magnetic shock models, they showed   that the dissipation of a small fraction ($<$10\%)  of the kinetic energy of the radio jet   heats the gas to a range of temperatures (Sect. \ref{tempmol}) and  is enough to  power the observed  mid-IR \hmol emission. 

An important difference between 4C12.50 and most MOHEGs is that it hosts a very luminous AGN (\citealt{Ogle2010}). \cite{Spoon2009b} and \cite{Guillard2012}  show that, while the lower ionization lines in the MIR are consistent with shocks, the high ionization lines (in particular [NeV]), arise primarily from photoionization of the gas by the AGN.  The presence of strong coronal lines (Sect. \ref{blend}) indicates that it must  contribute to the excitation of at least the ionized gas. This could explain why this system, which is expected to host strong jet-induced shocks affecting the molecular gas (see also previous section), is located in the AGN area of the diagnostic diagram.

\subsubsection{Kinematics: a  molecular outflow cannot be confirmed}
\label{kinmol}

 \cite{Dasyra2011} identified prominent blue wings in two out of the three H$_2$ MIR lines detected in the Spitzer spectrum of 4C12.50, \hmol(0-0) S(1) at 17.04 $\mu$m and S(2) at 12.28 $\mu$m. The main component, which they consider to trace the systemic velocity, is spectrally unresolved with FWHM$\la$550 km s$^{-1}$. The blue wing is $\sim$2.6 times fainter, it is shifted by  $\sim$-640 km s$^{-1}$ and has an instrumentally-corrected FWHM$\sim$521 km s$^{-1}$ (errors unavailable).  They propose this is produced by an AGN jet or wind-driven outflow. The  outflow  mass, 5.2$\times$10$^7$ M$_{\odot}$,  is a very high fraction, $\sim$27\%, of the total warm ($\sim$400 K) H$_2$  mass. \cite{Guillard2012} also  studied the MIR \hmol lines based on Spitzer data. They could not confirm the outflow and   report the tentative detection of a blue wing in \hmol(0-0) S(1) only.  

We investigate in this section whether the molecular outflow is detected in the NIR \hmol lines.

\begin{figure}
\centering
\includegraphics[width=0.5\textwidth]{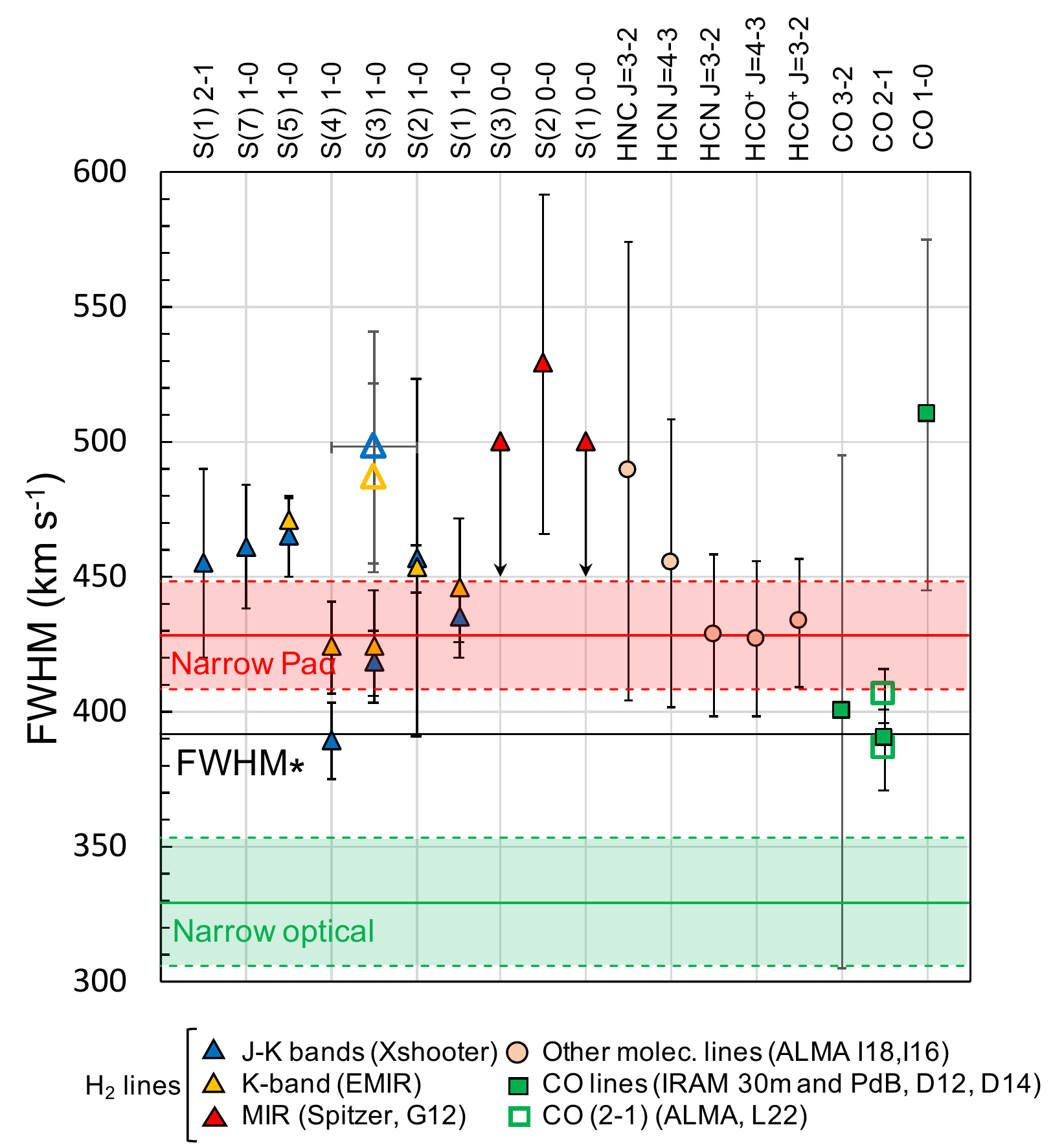}
\caption{Intrinsic FWHM  of molecular lines detected in 4C12.50.   The solid and open triangles for  S(3) (1-0) are the values from Tables \ref{tablelines}  and \ref{fith2sivi}  respectively. The two green open squares are ALMA CO(2-1)  measurements for the nuclear (top, 0.1$\arcsec$ radius aperture) and integrated (bottom, 1.5$\arcsec$ radius aperture) spectra (\citealt{Lamperti2022}). The solid  horizontal black line marks  FWHM$_{\rm *}$=392 \kms (\citealt{Dasyra2006}).  The  coloured areas correspond to the FWHM ($\pm$errors) of the narrow (systemic)  component of \pa (red) and the optical (green) ionized gas lines. D12: \citealt{Dasyra2012}; D14: \citealt{Dasyra2014}; G12: \citealt{Guillard2012}; I16: \citealt{Imanishi2016}; I18: \citealt{Imanishi2018}.}
\label{compkinem}
\end{figure}

\begin{figure*}
\includegraphics[width=1\textwidth]{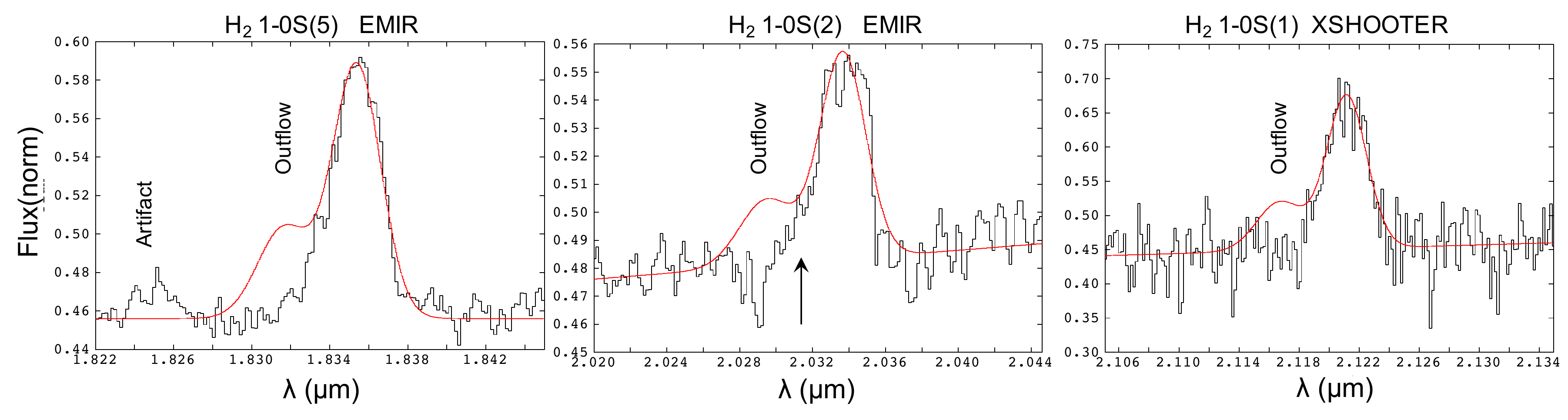}
\caption{Observed emission (black) of \hmol S(5) (left), S(2) (middle) and S(1) (right) and the expected profiles (red), for which we assume  the presence of a molecular outflow with similar kinematic properties and relative flux contribution as in the MIR \hmol lines. The troughs at the peak of the S(2) and S(1) are due to the noise.}
\label{split}
\end{figure*}

The hot molecular gas has much simpler kinematics than the ionized gas, whose motions are strongly defined by the   outflow triggered by the radio source (Sect. \ref{secpa}).    The NIR  \hmol line profiles are well reproduced by single Gaussians  and have instrumentally-corrected FWHM with median values $\sim$446$\pm$13  and 455$\pm$10   \kms for the EMIR and Xshooter spectra respectively (Table \ref{tablelines}). No broad components that may be indicative of an outflow are detected for any line (see below).

As shown in Fig. \ref{compkinem}, the lines emitted by \hmol and other molecular species in 4C12.50 have  FWHM$\ga$400 km s$^{-1}$  (Fig. \ref{compkinem}), although some are affected by large uncertainties\footnote{\hmol   0-0 S(0) at 28.22$\mu$m is surprisingly broad FWHM=906$\pm$91 \kmsb  (\citealt{Guillard2012}). Because this line   is in a noisy part of the Spitzer spectrum (see their Fig. 2) and given the inconsistency of its FWHM in comparison with all other molecular lines, we will not consider it.}. The CO(1-0)  and \hmol NIR lines are rather broad compared with different types of galaxies at $z<$0.5, but not extreme if U/LIRGs and active galaxies, including quasars, are  considered (\citealt{Murphy2001,Ardila2004,Ardila2005,Colina2005,Piqueras2012,Riffel2013,Villar2013,Cortzen2019,Ramos2022,Lamperti2022}).

Information on the potential presence of gas turbulence can be obtained by comparing with the stellar FWHM$_{\rm *}$. As in many interacting systems, complex, non-ordered stellar motions have been identified in 4C12.50 (\citealt{Perna2021}). 
 \cite{Dasyra2006} measured FWHM$_{\rm *}$=392$\pm$112 km s$^{-1}$, which is affected by a large uncertainty.    The narrow (systemic) component of the  emission lines from the ionized gas  has FWHM=340$\pm$23 \kms for the optical lines (\citealt{Holt2003}; 319$\pm$6 \kms according to \citealt{Rose2018}) and  FWHM=419$\pm$18 \kms or  392$\pm$15 \kms for \pa according to the EMIR and the X-shooter spectra respectively; Table \ref{tablelines}). All are consistent with FWHM$_{\rm *}$ within the errors, although the FWHM of the narrow  \pa  is somewhat broader (3.2$\sigma$ significance) in comparison with the optical lines.  This is not surprising, given the rich dust content of 4C12.50. Slightly larger stellar velocity widths are often inferred for galaxies using  NIR stellar features compared with the optical values. This  suggests that the NIR features probe more deeply embedded (and therefore higher velocity dispersion) stellar populations than the optical ones (\citealt{Caglar2020}).

 Depending on which value we use (Fig. \ref{compkinem}),  the molecular lines are all broader than the narrow optical component or  similar to the narrow Pa$\alpha$.   This is  not likely to be affected by aperture effects, given the diversity of aperture sizes for the data and the consistency of the result for most lines.  Based on this comparison, therefore, it is not possible to confirm whether the hot (neither the warm) \hmol gas  shows turbulent motions in relation to the systemic motions. 

An indication of turbulence of the warm and hot molecular gas is suggested by the fact that all lines appear to be  broader than CO(2-1) and possibily CO(1-0). The origin is however unknown.
 Although the dominant kinematic component of the hot molecular gas in ULIRGs and Seyferts is rotation,  non rotational components are often also identified  (e.g. \citealt{Bianchin2021}).    They could be gas elements  out of dynamical equilibrium, such as gas streams related to  galaxy interactions,  inflows or outflows  \cite{Dasyra2011,Guillard2012,Fotopoulou2019}.

Another result of the kinematic analysis is that  a counterpart of the prominent MIR \hmol molecular outflow  identified by \cite{Dasyra2011} is not confirmed. We show next that  if the NIR \hmol lines had the same relative contribution of the outflow as in the MIR,  we should have detected it.

For this, we have created the expected spectral profiles of \hmol S(5), S(2) and S(1)   assuming the same kinematic substructure in \kms as the MIR lines (instrumental broadening has been taken into account). We then compared these   with the data. For S(1) we have used the Xshooter spectrum because the line is on the edge of the EMIR data.  The results are shown in Fig. \ref{split}. The red lines are the expected line profiles. The molecular outflow should have been detected for the three lines as a clear blue excess, but this is not the case.  

A faint blue excess may be hinted on the blue wing of S(2) in the EMIR spectrum (Fig. \ref{split}). However, it is not clear this is real based on the non detection in the Xshooter spectrum, the structure of the noise in adjacent spectral regions and the absence of the wing in other \hmol lines, both in the Xshooter  and EMIR spectra.

In summary, we find no  evidence for a hot molecular outflow in the NIR \hmol lines.   Kinematic turbulence is suggested by somewhat broader line widths in comparison with   CO(2-1) and CO(1-0). The origin of this turbulence can be diverse. Given the clear role of the  jet induced shocks  in heating the \hmol gas, it seems natural that they may also affect the kinematics inducing some turbulence   (see also\citealt{Guillard2012}).

It is possible that faint spectroscopic features related to the feedback induced by the jet are  lost in the overwhelming glare of the bright nuclear line emission in the spatially integrated spectra analyzed in this work.  NIR and MIR integral field spectroscopy at very high spatial resolution (for instance with NIRSPEC and/or MIRI on the JWST) would be of key value to map in two spatial dimensions the impact of the interaction between  the radio jet and the ambient hot and warm molecular gas with spatial resolution of FWHM$\sim$several$\times$100 pc.

\section{Discussion}
\label{discussion}

Observations of ionized and neutral gas outflows in radio galaxies    suggest that AGN
radio jet feedback has the potential to affect the   gaseous environment of their hosts from  nuclear to galactic scales,  and   out into the circumgalactic medium. This feedback mechanism may  also be relevant  in systems hosting   moderate power radio sources (e.g. Villar Mart\'\i n et al. \citeyear{Villar2017,Villar2021}, \citealt{Jarvis2019,Girdhar2022}). To determine whether and how the radio sources can regulate the star formation in their host galaxies it is necessary to understand how the molecular gas is affected (e.g. \citealt{Tadhunter2014,Morganti2021}). 

 4C12.50 is a very relevant system in this context. If radio induced feedback can regulate the star formation activity in galaxies, it is a promising candidate to reveal this phenomenon in action.The compact twin jet  is still within the region where a huge accumulation of molecular gas and dust formed during the course of a gas-rich merger, prior to  the coalescence with the companion secondary nucleus. This process has favoured intense star formation (see Sect. \ref{intro}).  The large mechanical energy of the powerful radio source and the high concentration of  gas have resulted on a strong jet-gas interaction that has triggered kinematically extreme ionized and neutral outflows (Sect. \ref{sec-ion} and references therein).

The regulation of the system's evolution by the jet may occur, on one hand, by expelling the dusty nuclear cocoon. The system may transition in this way from a ULIRG to an optically less obscured radio galaxy (Sect. \ref{intro}).  On the other hand, the radio source could act on the molecular reservoir by heating (or cooling) gas, which may result in quenching (or triggering) star formation and thus halting (or triggering) the growth of the galaxy.

\subsection{Is the radio jet clearing out  the cocoon?}
  Like many local ULIRGs, 4C12.50 hosts an obscured and very compact  nucleus.  Based on the CO(2-1) ALMA data described in \cite{Lamperti2022}, we measure   a deconvolved half light radius of the CO(2-1) emisssion of  $r_{\rm CO}=$358$\pm$2 pc (see also \citealt{Evans2002}). A molecular gas mass log($M_{\rm H2}$) = 9.36$\pm$0.05  \msun ($\sim$23\% of the total cold molecular gas mass) is enclosed within.  Most of the remaining cold molecular gas is in a $\sim$4 kpc wide disk (\citealt{Fotopoulou2019}).

 For comparison, compact  obscured nuclei with $r_{\rm CO}\sim$200-400 pc  and $M_{\rm H2}\sim$(0.1-several)$\times$10$^9$ \msun are common in local ULIRGs (\citealt{Condon1991,Soifer2000,Pereira2021}). The small sizes of the twin jet in 4C12.50 (total size $\sim$220 pc) and of the  ionized (radial size $r\sim$69 pc, \citealt{Tadhunter2018}) and neutral ($r\sim$100 pc, \citealt{Morganti2013}) outflow triggered by it   indicate that they   are well within the dusty cocoon. Therefore, the clearing up process may  be at work. 

Currently, the radio plasma  appears to be dragging a small fraction of the total cocoon mass,  with $\sim$8$\times$10$^5$ \msun of ionized gas and 1.6$\times$10$^4$ \msun  of  neutral gas at a total mass outflow rate of $\sim$10 \msun yr$^{-1}$ at most (\citealt{Holt2011,Morganti2013,Rose2018}).   
Considering all studies of the molecular gas component, there is no solid evidence for a  molecular outflow in 4C12.50, neither in emission, nor in absorption (Sect. \ref{intro} and \ref{kinmol}), not at any temperature, hot (this work), warm or cold.
In spite of the clear and dramatic impact of the jet  on  the ionized and neutral gas kinematics, it  has no obvious effects on the kinematics of the hot (neither warm) molecular gas except, possibly, some enhanced turbulence in comparison with the cold molecular gas. This is not enough to remove significant amounts of molecular gas and clear out the central cocoon   (\citealt{Guillard2012}). As discussed by the later authors,  this implies that dynamical coupling between the
molecular gas and the  ionized and neutral outflowing gas is weak.

At the total observed $\dot{M}$, and assuming the unlikely case that all the gas is successfully removed from the central region, the radio source would have to be trapped and removing the gas within the cocoon for an unrealistically long time $\sim$2.3$\times$10$^8$ yr. This is as long as the maximum  radio source dynamical ages in radio galaxies in general. These are typically $\la$several$\times$10$^8$ yr, with the longest ages measured in giant radio galaxies (e.g. \citealt{Machalski2007,Machalski2009}).

At the moment, it appears that the radio jet is displacing mass at a too slow rate to efficiently clear up the dusty cocoon. It is unclear whether the removal of a significantly lower amount of mass would suffice to promote the transtion into an optically less obscured radio galaxy.

\subsection{Has star formation been suppressed by jet-induced feedback in 4C12.50?}

It is not clear whether the outflow can affect substantially the star formation activity in 4C12.50, given the moderate mass outflow rate ($\dot{M}<  SFR$), kinetic power  and small volume (see \citealt{Rose2018} for a detailed dicussion).  The radio source size is   $\sim$220 pc.  The affected volume is much smaller than the galactic bulge (the effective radius of the galaxy hosting the western nucleus is $r_{\rm eff}=$2.59$\pm$0.58 kpc, \citealt{Dasyra2006}).  Star formation   may be affected within the inner $r\la$100 pc,  the maximum estimated outflow size, but the  impact on larger scales is lacking evidence.     As argued by \cite{Rose2018}, although the presence of a lower density outflow component that has a high mass and contributes relatively little to the emission line fluxes cannot be ruled out, there is currently no  observational evidence for it.

Even if a powerful molecular outflow is not triggered, supression of star formation may  occur as a consequence of other processes related to the jet,    such as  molecular gas heating and/or the injection of turbulence. Evidence for this process  lies in the  unusually high temperature ($T_{\rm rot}$=3020$\pm$160 K) of the hot component, the power law $T$-mass function relation  (Sect. \ref{tempmol} and \ref{massmol}) and the fact that shocks are needed to explain the MIR \hmol emission (Sect. \ref{excitmol}; \citealt{Ogle2010,Guillard2012}).  Shocks can be generated in different ways. In ULIRGs in particular, interactions with nearby galaxies can excite large-scale shocks that will cool by means of \hmol emission (\citealt{Zakamska2010,Rich2015}).   In 4C12.50,  a natural scenario is that the  compact, powerful jet plays a major role on this regard.    The clear impact   of the interaction with the ionized and neutral components suggests that shocks are indeed present. If the jet encounters molecular gas on its path, it may  inject mechanical energy capable of heating it, even if the kinematics is not significantly affected. 

We investigate next whether there is evidence for star formation supression in 4C12.50.
\cite{Lanz2016} found that MOHEGs  (including 4C12.50) fall below the K-S relation of galaxies, log($\Sigma_{\rm SFR}$) vs. log($\Sigma_{\rm CO}$) (\citealt{Kennicutt1998}). These are the surface density of star formation  and   the surface density of molecular gas respectively. According to the authors, this suggests that   the star formation rate (SFR) is suppressed by a factor of 3-6, depending on how the molecular gas mass is estimated. Approximately 25\% of their sample shows  a suppression by more than a factor of 10. For 4C12.50, they found a factor  $\sim$10 in comparison with normal galaxies and a factor $\sim$100  in comparison with ULIRGs.  They suggested that  the shocks driven by the radio jets are responsible for the suppression, by injecting turbulence into the interstellar medium
(ISM).  They also found that the degree of SFR suppression does not correlate with indicators of jet feedback including jet power, diffuse X-ray emission, or intensity of warm molecular \hmol emission.

Using the ALMA data  previously mentioned, we have revised the location of 4C12.50 in the K–S relation   of different galaxy types. For the purpose of comparison with \cite{Lanz2016}, we  calculate $\Sigma_{\rm SFR}=\frac{SFR}{\pi r_{\rm SF}^2}$ and    $\Sigma_{\rm M_{\rm H_2}}=\frac{M_{\rm H_2}}{\pi r_{\rm H_2}^2}$,  where $r_{\rm SF}$ and $r_{\rm H_2}$ are the radial sizes  of the star forming region and of the molecular gas distribution respectively. 

The main sources of uncertainty  to determine both $\Sigma$ values  come from  the  uncertain SFR and the areas of the star forming region and the molecular gas distributions.

The main difficulty to estimate the  SFR is the uncertain fraction of AGN contribution to  $L_{\rm IR}$, $f_{\rm IR}=100 \times \frac{L_{\rm IR}^{\rm AGN}}{L_{\rm IR}}$. \cite{Veilleux2009} applied six different methods to infer the fraction of AGN contribution to the bolometric luminosity $L_{\rm bol}$, $f_{\rm bol}=100 \times \frac{L_{\rm bol}^{\rm AGN}}{L_{\rm bol}}$ and obtained values in the range $\sim$28-84\%, with an average of 57 \%. \cite{Perna2021} constrained the range further to $f_{\rm bol}\sim$60-82\% depending on the method. Since $\sim$90\% of $L_{\rm bol}$ is emitted in the infrared for 4C12.50, we assume $f_{\rm bol}\sim f_{\rm IR}$. $f_{\rm IR}\sim$60-82\% is a reasonable range of values also  based on the the ratio of the mid to far infrared continuum fluxes of 4C12.50, log($\frac{F_{5-25 \mu m}}{{\rm 40-122 \mu m}}$)=-0.19, which is intermediate between  starburst dominated ULIRGs ($\sim$-1.25) and AGN  dominated ULIRGs ($\sim$0.35) (\citealt{Veilleux2009}).  This suggestas a significant but not total contamination of a MIR bump due to the AGN (\citealt{Lanz2016}).

This implies  SFR$\sim$63-141 \msun yr$^{-1}$  (\citealt{Kennicutt1998}). For comparison  \cite{Rupke2005b} quote 101 \msun yr$^{-1}$. These SFR are in  the range of other ULIRGs (e.g. \citealt{Daddi2010a,DeLooze2014,Perna2021}),  SFR$\ga$100 \msun yr$^{-1}$ is consistent with the value expected for its gas mass   (see Fig. 1 in \citealt{Daddi2010a}).
\citealt{Lanz2016} fitted the spectral energy distribution (SED) of 4C12.50 with both an AGN and a SB component, which is essential to obtain a more accurate  $f_{\rm IR}$. They inferred SFR$\sim$24 \msun yr$^{-1}$ for 4C12.50 and, thus,   $f_{\rm IR}\sim$93\%. This is indeed bellow the value expected for its gas mass. Notice, however, that the authors warn that a combination of dust temperatures not implemented in their method would be more adequate to fit the IR SED.  
Given all uncertainties, we will consider three possibilities: SFR=24, 63 and 141 \msun yr$^{-1}$ .
 
Regarding the areas of the star forming region and the molecular gas distributions, we have considered several possibilities  summarized in Table \ref{supr}.
 We measure a half light radius  of the CO(2-1) emitting region $r_{\rm CO}$=358$\pm$2 pc (see previous section), which  is very similar  to  the median value, 320 pc, inferred by \cite{Pereira2021} for their  sample of nearby ULIRGs.  We  consider  this $r_{\rm CO}$ a reasonable upper limit for $r_{\rm SF}$, since $r_{\rm SF}<r_{\rm CO}$ invariably in ULIRGs (\citealt{Pereira2021}). 
This is also consistent  with  MIR observations which indicate that most of the star formation in ULIRGs occurs in  very compact regions ($<$1 kpc; e.g. \citealt{Soifer2000,Diaz2010,Alonso2016}).  his implies  lower  limits for log($\Sigma_{\rm SF}$) of $\ga$1.76 or $\ga$2.19 and 2.54 depending of the three assumed SFR values for 4C12.50. For comparison, most ULIRGs  in \cite{Pereira2021} sample are in the range 2.8-4.3 \footnote{For the purpose of this comparison, we have multiplied their values by 2, since they estimate the area as $A=2 \pi r^2$.}.  

We have considered different possible values for $r_{\rm H2}$. In case (A)  (Table \ref{supr}), $r_{\rm H2}$= $r_{\rm CO}$ and the mass within is log($M_{\rm H_2}$/M$_{\rm \odot}$)=9.36.  This implies log($\Sigma_{\rm M_{H_2}}$=3.76). This is within the range  2.59-4.49 (log of the median  3.91) in \cite{Pereira2021}   (see also \citealt{Bellocchi2022}). 

 Based on the CO(1-0), (3-2), and (4-3) line obsevations with ALMA), \cite{Fotopoulou2019} discovered that most of the total  molecular gas of 4C12.50 is within  $r\sim$2 kpc (this is consistent with our CO(2-1) ALMA data), including   a  disk of radius $r_{\rm disk}$=2 kpc, associated with the western nucleus. Although they detected more extended molecular gas, its mass is relatively low. In case (B), we thus assume $r_{\rm H_2}$=2 kpc and log($M_{\rm H_2}$/M$_{\rm \odot}$)=10.00, which is the total cold molecular gas of the system. For the sake of comparison with \cite{Lanz2016}, we assume  $r_{\rm SF}$=$r_{\rm H_2}$, although this $r_{\rm SF}$ is  likely to be unrealistically large (see above).  

Finally, we will assume  the values adopted by \cite{Lanz2016}, $r_{\rm SF}$=$r_{\rm H_2}$=4.2 kpc (case C). This  is the extent of the main CO(1-0) spatial component, which  \cite{Dasyra2014} found to be marginally resolved within the beam size based on IRAM Plateau de Bure Interferometer data.  We emphasize that these authors used $\alpha_{\rm CO}$=4.3 (K km$^{-1}$ s$^{-1}$ pc$^2$)$^{-1}$ (instead of 0.78), leading to a significantly higher total molecular gas mass of log($M_{\rm H_2}$/M$_{\rm \odot}$)=10.73. 

\begin{figure}
\centering
\includegraphics[width=0.48\textwidth]{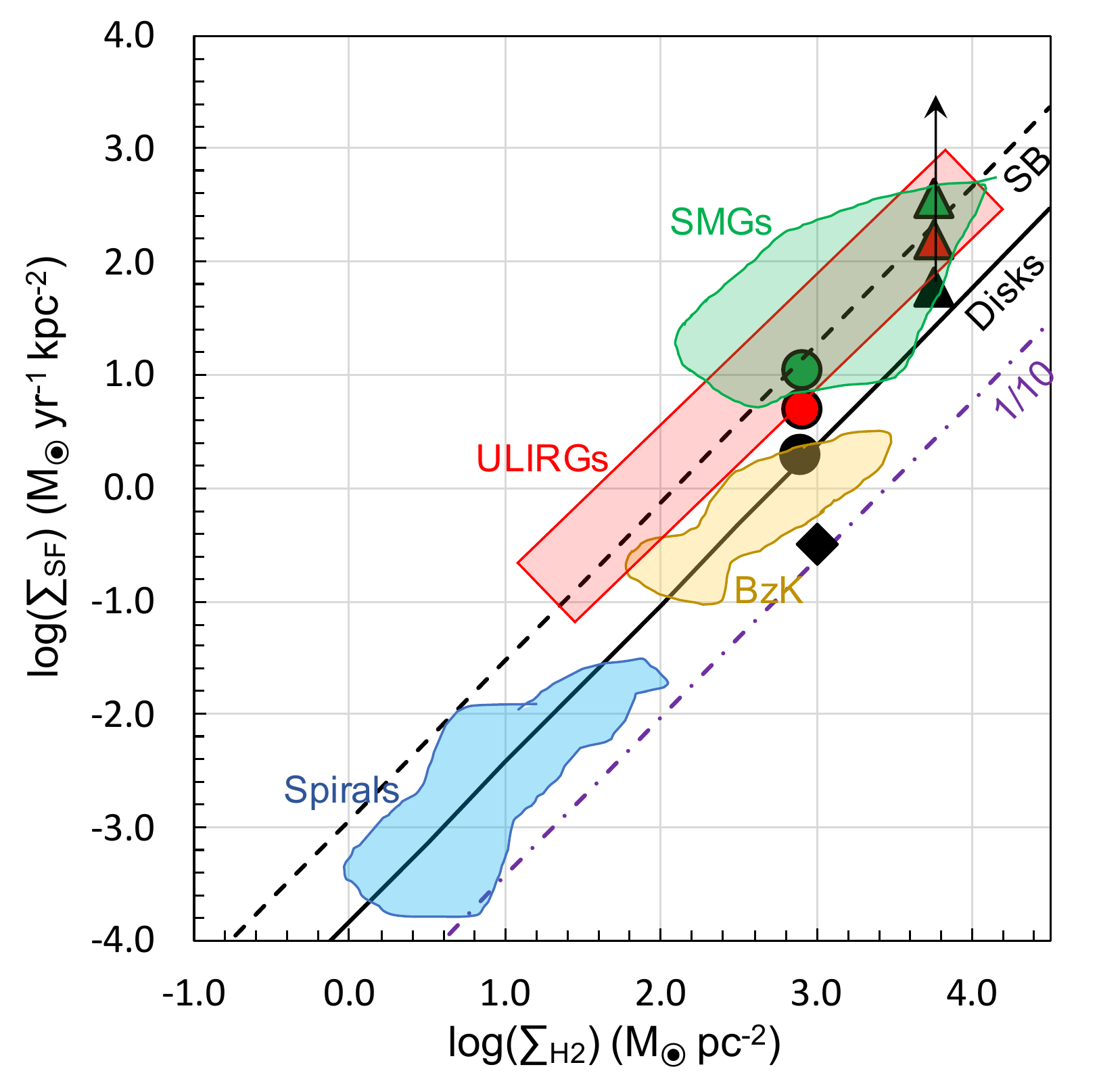}
\caption{Surface density of star formation compared to gas (cold molecular + neutral)  density gas (K–S diagrams; \citealt{Kennicutt1998}). This figure has been adapted from Fig. 2 in \cite{Daddi2010a}. The coloured areas roughly enclose different galaxy types indicated with labels. The black solid line ("disks sequence") is a fit to local spirals and $z\sim$1.5 BzK selected galaxies (slope of 1.42). The upper dashed line is the same relation shifted up by 0.9 dex to fit local (U)LIRGs and SMGs.  The dot-dashed purple  line corresponds to a star formation suppression of a factor of 10 relative to the disks sequence.  The coloured symbols are different locations of 4C12.50 for the diversity of parameters  used in our calculations. The triangles correspond to  cases (A) and the circles correspond to cases  (B) in Table \ref{supr}).  The black diamond comes from \citealt{Lanz2016} (case (C)). The triangles  represent the values obtained from recent high angular resolution ALMA CO(2-1) data. }
\label{ks}
\end{figure}

  $\Sigma_{\rm M_{H_2}}$ and  $\Sigma_{\rm SFR}$  for cases (A) to (C) are shown in Table \ref{supr} and their location  on the the K-S relation in Fig. \ref{ks}. As argued by \cite{Daddi2010a}, this diagram suggests the
existence of two different SF regimes: a long-lasting mode for disks and a more rapid mode for starbursts (including U/LIRGs and submillimeter galaxies (SMG)), the latter probably occurring during major mergers or in dense nuclear SF regions. They considered the total neutral and molecular gas. We use only the cold molecular gas in 4C12.50, because the neutral gas  mass in ULIRGs is in likely negligible  in comparison  (\citealt{Daddi2010b}). 

Being a ULIRG undergoing a major merger, we would expect 4C12.50 to lie  near  the starburst sequence.  On the contrary, \cite{Lanz2016} found that the object is not only well below  it, but well below the "normal'' disks sequence as well (see black diamond in Fig. \ref{ks}). 

Based on our revised location of 4C12.50 in the K-S diagram we 
cannot confirm this conclusion. In fact, the location based on the new  ALMA measurements (Case (A) lower limits represented as triangles in the figure) in Table \ref{supr}),   all lie close or above the starburst region.  Given that $r_{\rm CO}$ is most probably a lower limit of $r_{\rm SF}$, we can  claim that we find no evidence for star formation supression in this object. Positive feedback, in the sense of an enhacement of the star formation activity induced by the jet, may be present. To investigate this    more accurate determinations of both the SFR and, specially,  $r_{\rm SF}$ would be very valuable.

The main reason for the discrepancy with \cite{Lanz2016} is their assumption of  unrealistically large $r_{\rm H2} \sim r_{\rm SF}\sim$4.2 kpc. Another important difference is the lower SFR$\sim$24 \msun yr$^{-1}$ and the assumed $\alpha_{\rm CO}$=4.3. While this  conversion factor is appropriate for galaxies in the disk sequence, $\alpha_{\rm CO}$=0.78 is more appropriate for U/LIRGs. This difference was taken into account by \cite{Daddi2010a}, who used $\alpha_{\rm CO}$ values adequate for each galaxy type.

\begin{table*}
\centering
\begin{tabular}{lccccccl}
\hline
Case & $r_{\rm H_2}$  & $r_{\rm SF}$ & log($M_{\rm H_2}^{enc}$) & SFR &  log($\Sigma_{\rm M_{H_2}}$) & log($\Sigma_{\rm SFR}$) & Symbol \\ 
 &	 kpc &	 kpc & \msun & \msun yr$^{-1}$  & \msun pc$^{-2}$ & \msun yr$^{-}$ kpc$^{-2}$ & \\
\hline
  &  & \multicolumn{3}{c}{(A) $r_{\rm H_2}$=$r_{\rm SF}$=$r_{\rm CO}$, $\alpha_{\rm CO}$=0.78 }&    \\
A1 &		0.358 & 	0.358 &   9.36 &  141 & 3.76 & 2.54  & Green triangle \\
A2 & 	0.358 &	0.358 & 9.36	 &   63 & 3.76 & 2.19 & Red triangle\\
A3 &	0.358 &	0.358 & 9.36	 &   24 & 3.76 & 1.76 & Black triangle\\ 
 &  & \multicolumn{3}{c}{(B) $r_{\rm H_2}$=$r_{\rm SF}$=$r_{\rm disk}$, $\alpha_{\rm CO}$=0.78 }&    \\
B1 &	2.0 &	2.0 & 10.00	&   141  & 2.90 & 1.05  & Green circle\\
B2 &	2.0 &	2.0 & 10.00	&   63  &  2.90 & 0.70  & Red circle \\
B3 &	2.0 &	2.0 & 10.00	&   24  & 2.90 & 0.28  & Black circle\\
&  & \multicolumn{3}{c}{(C) $r_{\rm H_2}$=$r_{\rm SF}$=4.2 kpc, $\alpha_{\rm CO}$=4.3}&    \\
C & 4.2	 &  4.2	 &  10.73 & 24 & 3.03 & -0.36  & Black  diamond\\ \hline
\end{tabular}	
\caption{Surface density of star formation and of cold molecular gas for different assumptions of SFR and radial sizes of the SF ($r_{\rm SF}$) and molecular gas ($r_{\rm H_2}$) distributions.  For the purpose of comparison with Lanz et al. (\citeyear{Lanz2016}), we have assumed the area corresping to each radial size, $r$,  as $A = \pi r^2$.    $M_{\rm H_2}^{enc}$ is the assumed cold molecular gas mass enclosed within $r_{\rm H_2}$.  SFR in units of \msun yr$^{-1}$. $\Sigma_{\rm M_{H_2}}$ in  \msun pc$^{-2}$. $r_{\rm CO}$=358 pc  is the CO(2-1)  half light radius inferred with ALMA data (see text). $r_{\rm disk}$=2 kpc is the radius of the molecular disk discovered by Fotopoulou et al. (\citeyear{Fotopoulou2019}). Case (C) corresponds to  Lanz et al. (\citeyear{Lanz2016}). All molecular masses were calculated with $\alpha_{\rm CO}$=0.78, except  in case (C), for which $\alpha_{\rm CO}$=4.3 as in \cite{Lanz2016}. The last column specifies the symbol used in Fig. \ref{ks}.} 
\label{supr}
\end{table*}

\subsection{No solid evidence for a significant impact of radio induced feedback on 4C12.50 evolution}

Overall, there is no solid evidence for the radio jet to have a significant impact on the evolution of 4C12.50, not by clearing out the dusty central cocoon, nor by suppressing the star formation activity.

It is possible that we are observing 4C12.50 in a very early phase of the radio activity. 
At the moment, the radio source appears to be affecting just a small volume.   However, as   times passes, unless its advance is frustrated by the rich gaseous medium the jet will propagate through the galaxy  and possibly expand beyond it at some point.  Signs of a previous radio outburst stretching outside the host   shows that this has happened in the past (\citealt{Lister2003,Stanghellini2005}). If the radio jet decelerates and inflates into large expanding bubbles, it will affect a much larger galactic volume. Whether  it will carry enough energy to expel the cocoon and/or to prevent
further gas cooling and/or to inject turbulence, affecting in this way future star formation is unknown (\citealt{Mukherjee2016,Morganti2021}).

The presence of other feedback mechanisms acting on different spatial scales  should not be forgotten  (AGN wind and/or starburst winds).  Their action is suggested by 
\cite{Rupke2005a} results. They detected blueshifted NaID absorption due to a neutral outflow  in the Western nucleus with $V_ {\rm max}$= 364 \kmsb, $M\sim 8\times10^8$ \msunb, $\dot M =0.88$ \msun yr$^{-1}$ and log($\dot E \rm (erg~s^{-1})$)$\sim$41.16 or $\sim$0.001\% of $L_{\rm bol}$ (but see \citealt{Perna2021}). They locate it up to a radius of $\sim$10-15 kpc and at a large angle from  the radio axis and thus, there is no clear link between this outflow and the radio source.

\section{Summary and conclusions}

We present for the first time a detailed analysis of  the hot ($>$1500 K) molecular \hmol gas  in the ultraluminous-infrared radio galaxy  4C12.50 at $z$=0.122 based on GTC EMIR  and VLT  Xshooter near-infarred long slit spectroscopy of the western primary nucleus. New results on the ionized (including coronal) phase are also presented.

 $\bullet$ 4C12.50 hosts a large  hot molecular gas content, with $M_{\rm H_2}^{\rm hot}$=(2.10$\pm$0.44)$\times$10$^{4}$ M$_{\rm \odot}$. This is consistent with its high infrared luminosity (in the ULIRG regime) and with the large mass of cold ($\la$25 K) molecular gas.

$\bullet$ An unusually high rotational temperature $T_{\rm rot}$ =3020$\pm$160 K is inferred for H$_2$, that is at the high end of values measured in galaxies in general.

$\bullet$ The molecular gas mass obeys  a power law temperature distribution   $\frac{d M_{\rm H2}}{dT}\propto T^{-5}$   from T$\sim$300 K and up to  $\sim$3000 K.  This is consistent with paraboloidal bow shock model predictions. This, and the high $T_{\rm rot}$, suggests that, as found by other authors for the warm \hmol MIR emission lines, shocks (probably induced by the radio jet) contribute to the heating and excitation of the hot molecular gas. The jet induced shocks can heat the molecular gas, even it the dynamical coupling is weak.

$\bullet$ The \hmol 2-0 S(3)$\lambda$1.1175 / 2-1 S(3)$\lambda$2.0735 ratio implies an extinction of 
$E_{\rm B-V}^{\rm mol}$=0.35$\pm$0.08 for the hot molecular gas.  For comparison,  the extinction inferred by other authors for the integrated ionized gas emission is $E_{\rm B-V}^{\rm ion}$=0.59$\pm$0.11.

$\bullet$ We find no evidence for a hot molecular outflow in the NIR \hmol lines.   The warm ($\sim$400 K) \hmol outflow tentatively identified in the MIR lines by other authors should have been detected, if the relative flux contribution  was the same. This is not the case. 

$\bullet$ In spite of the dramatic impact of the radio jet induced shocks on the dynamics of the ionized and neutral phases of 4C12.50, the dynamics of the hot molecular gas   is not obviously affected. This is consistent with previous works that suggest the poor coupling of the outflowing ionized and neutral outflows with the molecular gas at lower temperatures.

$\bullet$ The extreme   Pa$\alpha$ kinematics are consistent with those of the optical lines, including several forbidden ones. The prominent broad blueshifted  \pa excess   is not produced by the broad line region. It is due to the kinematically extreme ionized outflow previously identified in the optical.   The outflow dominates (81$\pm$4\%) the Pa$\alpha$ flux.

$\bullet$  Most  of (possibily all) the  coronal gas appears to be outflowing, with blueshifted and very broad [SiVI]$\lambda$1.963 and [FeVII]$\lambda$6087  (FWHM$=$3430$\pm$450 \kms and   $\Delta V_{\rm sys}=$-1350$\pm$190 \kms for the broadest component of [SiVI]). 

$\bullet$ The relative contribution  of the most turbulent (broadest) outflowing gas component to the total forbidden line fluxes, $\frac{F_{\rm br}}{F_{\rm tot}}$, correlates with the critical density $n_{\rm crit}$ in log. This suggests that a single mechanism (the radio jet) is responsible for the outflow identified in all  lines emitted by the ionized phase, from the coronal to the lowest ionization.

$\bullet$ If the outflowing ionized gas moving away from the observer is completely extinguished, it could be trully kinematically  extreme, with FWHM$\sim$5400 \kms ($\sim$7000 \kms for the coronal gas), only comparable to those seen in a handful of high $z$ extremely luminous quasars. Its mass and energetics would still be moderate.

$\bullet$ Based on ALMA CO(2-1) data and a new estimation of the star forming rate, we revise the location of 4C12.50 in the K-S diagram. Contrary to other studies, we claim that there is  no evidence for star formation suppresion in this object. Positive feedback as a consequence of jet induced star formation is not discarded.

$\bullet$ If radio induced feedback can regulate the star formation activity in galaxies, 4C12.50 is a promising candidate to reveal this phenomenon in action. The rich amount of knowledge available for this object, however, has not provided solid evidence so far for this to be the case.   We find   no solid evidence for current or past impact of this mechanism on the evolution of this system, neither by clearing out the dusty central cocoon efficiently, nor by suppressing the star formation activity.

\section*{Acknowledgments}

We thank an anonymous referee for revising the paper and contributing with constructive comments.
This research has made use of grants  PGC2018-094671-BI00, PID2021-124665NB-I00 (MVM, AC and AAH) and   PIB2021-127718NB-100 (LC)  by the Spanish Ministry of Science and Innovation/State Agency of Research MCIN/AEI/ 10.13039/501100011033 and by "ERDF A way of making Europe". IL acknowledges support from the Spanish Ministry of Science and Innovation (MCIN) by means of the Recovery and Resilience Facility, and the Agencia Estatal de Investigación (AEI) under the projects with references BDC20221289 and PID2019-105423GA-I00.  
EB acknowledges the María Zambrano program of the Spanish Ministerio de Universidades funded by the Next Generation European Union and is also partly supported by grant RTI2018-096188-B-I00 funded by MCIN/AEI/10.13039/501100011033.

Based on observations carried out at the Observatorio Roque de los Muchachos (La Palma, Spain) with EMIR on GTC (program GTC16-21B) and   at the European
Organization for Astronomical Research in the Southern hemisphere with Xshooter on VLT (ESO programme 091.B-0256(A)).
 We  thank the observatories staff for their support with the  observations.   It also makes use of  the following ALMA data: ADS/JAO.ALMA\#2018.1.00699.S. ALMA is a partnership of ESO (representing its member states), NSF (USA) and NINS (Japan), together with NRC (Canada) and NSC and ASIAA (Taiwan) and KASI (Republic of Korea), in cooperation with the Republic of Chile. The Joint ALMA Observatory is operated by ESO, AUI/NRAO and NAOJ. The National Radio Astronomy Observatory is a facility of the National Science Foundation operated under cooperative agreement by Associated Universities, Inc. 

 This research has made use of: 1) the VizieR catalogue access tool, CDS,
 Strasbourg, France. The original description of the VizieR service was
 published in \cite{Ochsenbein2000};   
2) the Cosmology calculator by \cite{Wright2006};
3) the NASA/IPAC Extragalactic Database (NED), which is operated by the Jet Propulsion Laboratory, California Institute of Technology, under contract with the National Aeronautics and Space Administration; 4) data from Sloan Digital Sky Survey. Funding for the SDSS and SDSS-II has been provided by the Alfred P. Sloan Foundation, the Participating Institutions, the National Science Foundation, the U.S. Department of Energy, the National Aeronautics and Space Administration, the Japanese Monbukagakusho, the Max Planck Society, and the Higher Education Funding Council for England. The SDSS Web Site is http://www.sdss.org/.


\begin{thebibliography}
\expandafter\ifx\csname natexlab\endcsname\relax\def\natexlab#1{#1}\fi


\bibitem[Alonso-Herrero et al.(2016)]{Alonso2016} Alonso-Herrero, A., Poulton, R., Roche, P.~F., et al.\ 2016, \mnras, 463, 2405

\bibitem[{\'A}lvarez-M{\'a}rquez et al.(2023)]{Alvarez2023} {\'A}lvarez-M{\'a}rquez, J., Labiano, A., Guillard, P., et al.\ 2023, A\&A, in press  (arXiv:2209.01695)


\bibitem[Batcheldor et al.(2007)]{Batcheldor2007} Batcheldor, D., Tadhunter, C., Holt, J., et al.\ 2007, \apj, 661, 70


\bibitem[Bellocchi et al.(2022)]{Bellocchi2022} Bellocchi, E., Pereira-Santaella, M., Colina, L., et al.\ 2022, \aap, 664, A60

\bibitem[Bennert et al.(2006)]{Bennert2006} Bennert, N., Jungwiert, B., Komossa, S., et al.\ 2006, \aap, 456, 953

\bibitem[Bianchin et al.(2021)]{Bianchin2021} Bianchin, M., Riffel, R.~A., Storchi-Bergmann, T., et al.\ 2021, \mnras. 

\bibitem[Bridges \& Irwin(1998)]{Bridges1998} Bridges, T.~J. \& Irwin, J.~A.\ 1998, \mnras, 300, 967

\bibitem[Caglar et al.(2020)]{Caglar2020} Caglar, T., Burtscher, L., Brandl, B., et al.\ 2020, \aap, 634, A114.

\bibitem[Calzetti et al.(2000)]{Calzetti2000} Calzetti, D., Armus, L., Bohlin, R.~C., et al.\ 2000, \apj, 533, 682

\bibitem[Cerqueira-Campos et al.(2021)]{Cerqueira2021} Cerqueira-Campos, F.~C., Rodr{\'\i}guez-Ardila, A., Riffel, R., et al.\ 2021, \mnras, 500, 2666.

\bibitem[Colina et al.(2005)]{Colina2005} Colina, L., Arribas, S., \& Monreal-Ibero, A.\ 2005, \apj, 621, 725. 

\bibitem[Colina et al.(2015)]{Colina2015} Colina, L., Piqueras L{\'o}pez, J., Arribas, S., et al.\ 2015, \aap, 578, A48

\bibitem[Condon et al.(1991)]{Condon1991} Condon, J.~J., Huang, Z.-P., Yin, Q.~F., et al.\ 1991, \apj, 378, 65

\bibitem[Cortzen et al.(2019)]{Cortzen2019} Cortzen, I., Garrett, J., Magdis, G., et al.\ 2019, \mnras, 482, 1618. 

\bibitem[Daddi et al.(2010a)]{Daddi2010a} Daddi, E., Elbaz, D., Walter, F., et al.\ 2010, \apjl, 714, L118

\bibitem[Daddi et al.(2010b)]{Daddi2010b} Daddi, E., Bournaud, F., Walter, F., et al.\ 2010, \apj, 713, 686


\bibitem[Dale et al.(2005)]{Dale2005} Dale, D.~A., Sheth, K., Helou, G., et al.\ 2005, \aj, 129, 2197

\bibitem[Dasyra et al.(2006)]{Dasyra2006} Dasyra, K.~M., Tacconi, L.~J., Davies, R.~I., et al.\ 2006, \apj, 638, 745

\bibitem[Dasyra \& Combes(2011)]{Dasyra2011} Dasyra, K.~M. \& Combes, F.\ 2011, \aap, 533, L10

\bibitem[Dasyra \& Combes(2012)]{Dasyra2012} Dasyra, K.~M. \& Combes, F.\ 2012, \aap, 541, L7

\bibitem[Dasyra et al.(2014)]{Dasyra2014} Dasyra, K.~M., Combes, F., Novak, G.~S., et al.\ 2014, \aap, 565, A46

\bibitem[Davies et al.(2003)]{Davies2003} Davies, R.~I., Sternberg, A., Lehnert, M., et al.\ 2003, \apj, 597, 907



\bibitem[De Looze et al.(2014)]{DeLooze2014} De Looze, I., Cormier, D., Lebouteiller, V., et al.\ 2014, \aap, 568, A62



\bibitem[De Robertis \& Osterbrock(1984)]{DeRobertis1984} De Robertis, M.~M. \& Osterbrock, D.~E.\ 1984, \apj, 286, 171

\bibitem[De Robertis \& Osterbrock(1986)]{DeRobertis1986} De Robertis, M.~M. \& Osterbrock, D.~E.\ 1986, \apj, 301, 727

\bibitem[den Brok et al.(2022)]{DenBrok2022} den Brok, J.~S., Koss, M.~J., Trakhtenbrot, B., et al.\ 2022, \apjs, 261, 7


\bibitem[D{\'\i}az-Santos et al.(2010)]{Diaz2010} D{\'\i}az-Santos, T., Charmandaris, V., Armus, L., et al.\ 2010, \apj, 723, 993

\bibitem[Emonts et al.(2014)]{Emonts2014} Emonts, B.~H.~C., Piqueras-L{\'o}pez, J., Colina, L., et al.\ 2014, \aap, 572, A40

\bibitem[Emonts et al.(2016)]{Emonts2016} Emonts, B.~H.~C., Morganti, R., Villar-Mart{\'\i}n, M., et al.\ 2016, \aap, 596, A19

\bibitem[Evans et al.(1999)]{Evans1999} Evans, A.~S., Kim, D.~C., Mazzarella, J.~M., et al.\ 1999, \apjl, 521, L107

\bibitem[Evans et al.(2002)]{Evans2002} Evans, A.~S., Mazzarella, J.~M., Surace, J.~A., et al.\ 2002, \apj, 580, 749

\bibitem[Farrah et al.(2022)]{Farrah2022} Farrah, D., Efstathiou, A., Afonso, J., et al.\ 2022, \mnras, 513, 4770

\bibitem[Fotopoulou et al.(2019)]{Fotopoulou2019} Fotopoulou, C.~M., Dasyra, K.~M., Combes, F., et al.\ 2019, \aap, 629, A30

\bibitem[Gilmore \& Shaw(1986)]{Gilmore1986} Gilmore, G. \& Shaw, M.~A.\ 1986, \nat, 321, 750

\bibitem[Grandi(1977)]{Grandi1977} Grandi, S.~A.\ 1977, \apj, 215, 446

\bibitem[Guillard et al.(2012)]{Guillard2012} Guillard, P., Ogle, P.~M., Emonts, B.~H.~C., et al.\ 2012, \apj, 747, 95

\bibitem[Girdhar et al.(2022)]{Girdhar2022} Girdhar, A., Harrison, C.~M., Mainieri, V., et al.\ 2022, \mnras, 512, 1608

\bibitem[Heckman et al.(1986)]{Heckman1986} Heckman, T.~M., Smith, E.~P., Baum, S.~A., et al.\ 1986, \apj, 311, 526

\bibitem[Holt et al.(2003)]{Holt2003} Holt, J., Tadhunter, C.~N., \& Morganti, R.\ 2003, \mnras, 342, 227

\bibitem[Holt et al.(2011)]{Holt2011} Holt, J., Tadhunter, C.~N., Morganti, R., et al.\ 2011, \mnras, 410, 1527

\bibitem[Imanishi et al.(2016)]{Imanishi2016} Imanishi, M., Nakanishi, K., \& Izumi, T.\ 2016, \aj, 152, 218

\bibitem[Imanishi et al.(2018)]{Imanishi2018} Imanishi, M., Nakanishi, K., \& Izumi, T.\ 2018, \apj, 856, 143

\bibitem[Jarvis et al.(2019)]{Jarvis2019} Jarvis, M.~E., Harrison, C.~M., Thomson, A.~P., et al.\ 2019, \mnras, 485, 2710

\bibitem[Kennicutt(1998)]{Kennicutt1998} Kennicutt, R.~C.\ 1998, \apj, 498, 541

\bibitem[Lanz et al.(2016)]{Lanz2016} Lanz, L., Ogle, P.~M., Alatalo, K., et al.\ 2016, \apj, 826, 29

  
\bibitem[Lamperti et al.(2017)]{Lamperti2017} Lamperti, I., Koss, M., Trakhtenbrot, B., et al.\ 2017, \mnras, 467, 540

\bibitem[Lamperti et al.(2022)]{Lamperti2022} Lamperti, I., et al. 2022, in press 

\bibitem[Larkin et al.(1998)]{Larkin1998} Larkin, J.~E., Armus, L., Knop, R.~A., et al.\ 1998, \apjs, 114, 59

\bibitem[Lister et al.(2003)]{Lister2003} Lister, M.~L., Kellermann, K.~I., Vermeulen, R.~C., et al.\ 2003, \apj, 584, 135. doi:10.1086/345666

\bibitem[Machalski et al.(2007)]{Machalski2007} Machalski, J., Chy{\.z}y, K.~T., Stawarz, {\L}., et al.\ 2007, \aap, 462, 43

\bibitem[Machalski et al.(2009)]{Machalski2009} Machalski, J., Jamrozy, M., \& Saikia, D.~J.\ 2009, \mnras, 395, 812

\bibitem[Ma{\l}ek et al.(2013)]{Malek2013} Ma{\l}ek, K., Pollo, A., Takeuchi, T.~T., et al.\ 2013, Earth, Planets and Space, 65, 1101

\bibitem[Mazzalay et al.(2013)]{Mazzalay2013} Mazzalay, X., Saglia, R.~P., Erwin, P., et al.\ 2013, \mnras, 428, 2389


\bibitem[Mezcua et al.(2015)]{Mezcua2015} Mezcua, M., Prieto, M.~A., Fern{\'a}ndez-Ontiveros, J.~A., et al.\ 2015, \mnras, 452, 4128

\bibitem[Mirabel(1989)]{Mirabel1989} Mirabel, I.~F.\ 1989, \apjl, 340, L13

\bibitem[Mirabel et al.(1989)]{Mirabel1989b} Mirabel, I.~F., Sanders, D.~B., \& Kazes, I.\ 1989, \apjl, 340, L9

\bibitem[Mittal et al.(2012)]{Mittal2012} Mittal, R., Oonk, J.~B.~R., Ferland, G.~J., et al.\ 2012, \mnras, 426, 2957

\bibitem[Morganti et al.(2003)]{Morganti2003} Morganti, R., Tadhunter, C.~N., Oosterloo, T.~A., et al.\ 2003, \pasa, 20, 129

\bibitem[Morganti et al.(2004)]{Morganti2004} Morganti, R., Oosterloo, T.~A., Tadhunter, C.~N., et al.\ 2004, \aap, 424, 119

\bibitem[Morganti et al.(2005)]{Morganti2005} Morganti, R., Tadhunter, C.~N., \& Oosterloo, T.~A.\ 2005, \aap, 444, L9

\bibitem[Morganti et al.(2013)]{Morganti2013} Morganti, R., Fogasy, J., Paragi, Z., et al.\ 2013, Science, 341, 1082

\bibitem[Morganti et al.(2021)]{Morganti2021} Morganti, R., Oosterloo, T., Tadhunter, C., et al.\ 2021, \aap, 656, A55

\bibitem[Mukherjee et al.(2016)]{Mukherjee2016} Mukherjee, D., Bicknell, G.~V., Sutherland, R., et al.\ 2016, \mnras, 461, 967

\bibitem[M{\"u}ller S{\'a}nchez et al.(2006)]{Muller2006} M{\"u}ller S{\'a}nchez, F., Davies, R.~I., Eisenhauer, F., et al.\ 2006, \aap, 454, 481



\bibitem[Murphy et al.(2001)]{Murphy2001} Murphy, T.~W., Soifer, B.~T., Matthews, K., et al.\ 2001, \aj, 121, 97

\bibitem[Neufeld \& Yuan(2008)]{Neufeld2008} Neufeld, D.~A. \& Yuan, Y.\ 2008, \apj, 678, 974

\bibitem[Oca{\~n}a Flaquer et al.(2010)]{Ocana2010} Oca{\~n}a Flaquer, B., Leon, S., Combes, F., et al.\ 2010, \aap, 518, A9

\bibitem[Ochsenbein et al.(2000)]{Ochsenbein2000} Ochsenbein, F., Bauer, P., \& Marcout, J.\ 2000, \aaps, 143, 23



\bibitem[O'Dea(1998)]{ODea1998} O'Dea, C.~P.\ 1998, \pasp, 110, 493. doi:10.1086/316162

\bibitem[O'Dea et al.(2000)]{Odea2000} O'Dea, C.~P., De Vries, W.~H., Worrall, D.~M., et al.\ 2000, \aj, 119, 478


\bibitem[O'Dea \& Saikia(2021)]{ODea2021} O'Dea, C.~P. \& Saikia, D.~J.\ 2021, \aapr, 29, 3

\bibitem[Ogle et al.(2010)]{Ogle2010} Ogle, P., Boulanger, F., Guillard, P., et al.\ 2010, \apj, 724, 1193

\bibitem[Oh et al.(2018)]{Oh2018} Oh, K., Koss, M., Markwardt, C.~B., et al.\ 2018, \apjs, 235, 4

\bibitem[Osterbrock \& Ferland(2006)]{Osterbrock2006} Osterbrock, D. E., \& Ferland, G. J. 2006, Astrophysics of gaseous nebulae and active galactic nuclei (University Science Books) 

\bibitem[Pereira-Santaella et al.(2014)]{Pereira2014} Pereira-Santaella, M., Spinoglio, L., van der Werf, P.~P., et al.\ 2014, \aap, 566, A49

\bibitem[Pereira-Santaella et al.(2021)]{Pereira2021} Pereira-Santaella, M., Colina, L., Garc{\'\i}a-Burillo, S., et al.\ 2021, \aap, 651, A42

\bibitem[Perna et al.(2021)]{Perna2021} Perna, M., Arribas, S., Pereira Santaella, M., et al.\ 2021, \aap, 646, A101

\bibitem[Perrotta et al.(2019)]{Perrotta2019} Perrotta, S., Hamann, F., Zakamska, N.~L., et al.\ 2019, \mnras, 488, 4126

\bibitem[Piqueras L{\'o}pez et al.(2012)]{Piqueras2012} Piqueras L{\'o}pez, J., Colina, L., Arribas, S., et al.\ 2012, \aap, 546, A64

\bibitem[Piqueras L{\'o}pez et al.(2013)]{Piqueras2013} Piqueras L{\'o}pez, J., Colina, L., Arribas, S., et al.\ 2013, \aap, 553, A85.

 \bibitem[\protect\citeauthoryear{Ramos Almeida et al.}{2009}]{Ramos2009} 
Ramos Almeida, C., P\'erez Garc\'\i a, \&  A. M., Acosta-Pulido, J., 2009, ApJ, 694, 1379

\bibitem[Ramos Almeida et al.(2019)]{Ramos2019} Ramos Almeida, C., Acosta-Pulido, J.~A., Tadhunter, C.~N., et al.\ 2019, \mnras, 487, L18

\bibitem[Ramos Almeida et al.(2022)]{Ramos2022} Ramos Almeida, C., Bischetti, M., Garc{\'\i}a-Burillo, S., et al.\ 2022, \aap, 658, A155. doi:10.1051/0004-6361/202141906

\bibitem[Rich et al.(2015)]{Rich2015} Rich, J.~A., Kewley, L.~J., \& Dopita, M.~A.\ 2015, \apjs, 221, 28

\bibitem[Riffel et al.(2006)]{Riffel2006} Riffel, R., Rodr{\'\i}guez-Ardila, A., \& Pastoriza, M.~G.\ 2006, \aap, 457, 61

\bibitem[Riffel et al.(2013)]{Riffel2013} Riffel, R.~A., Storchi-Bergmann, T., Riffel, R., et al.\ 2013, \mnras, 429, 2587

\bibitem[Riffel et al.(2014)]{Riffel2014} Riffel, R.~A., Vale, T.~B., Storchi-Bergmann, T., et al.\ 2014, \mnras, 442, 656

\bibitem[Riffel et al.(2021)]{Riffel2021} Riffel, R.~A., Storchi-Bergmann, T., Riffel, R., et al.\ 2021, \mnras, 504, 3265

\bibitem[Rodr{\'\i}guez-Ardila et al.(2002)]{Ardila2002} Rodr{\'\i}guez-Ardila, A., Viegas, S.~M., Pastoriza, M.~G., et al.\ 2002, \apj, 579, 214

\bibitem[Rodr{\'\i}guez-Ardila et al.(2004)]{Ardila2004} Rodr{\'\i}guez-Ardila, A., Pastoriza, M.~G., Viegas, S., et al.\ 2004, \aap, 425, 457

\bibitem[Rodr{\'\i}guez-Ardila et al.(2005)]{Ardila2005} Rodr{\'\i}guez-Ardila, A., Riffel, R., \& Pastoriza, M.~G.\ 2005, \mnras, 364, 1041

\bibitem[Rodr{\'\i}guez-Ardila et al.(2011)]{Ardila2011} Rodr{\'\i}guez-Ardila, A., Prieto, M.~A., Portilla, J.~G., et al.\ 2011, \apj, 743, 100.


\bibitem[Rodr{\'\i}guez Zaur{\'\i}n et al.(2013)]{Rodriguez2013} Rodr{\'\i}guez Zaur{\'\i}n, J., Tadhunter, C.~N., Rose, M., et al.\ 2013, \mnras, 432, 138

\bibitem[Rose et al.(2011)]{Rose2011} Rose, M., Tadhunter, C.~N., Holt, J., et al.\ 2011, \mnras, 414, 3360

\bibitem[Rose et al.(2018)]{Rose2018} Rose, M., Tadhunter, C., Ramos Almeida, C., et al.\ 2018, \mnras, 474, 128

\bibitem[Rupke et al.(2005a)]{Rupke2005a} Rupke, D.~S., Veilleux, S., \& Sanders, D.~B.\ 2005a, \apj, 632, 751


\bibitem[Rupke et al.(2005b)]{Rupke2005b} Rupke, D.~S., Veilleux, S., \& Sanders, D.~B.\ 2005b, \apjs, 632, 751



\bibitem[Sadler(2016)]{Sadler2016} Sadler, E.~M.\ 2016, Astronomische Nachrichten, 337, 105

\bibitem[Santoro et al.(2020)]{Santoro2020} Santoro, F., Tadhunter, C., Baron, D., et al.\ 2020, \aap, 644, A54

\bibitem[Scoville et al.(1982)]{Scoville1982} Scoville, N.~Z., Hall, D.~N.~B., Ridgway, S.~T., et al.\ 1982, \apj, 253, 136

\bibitem[Smol{\v{c}}i{\'c} \& Riechers(2011)]{Smolvic2011} Smol{\v{c}}i{\'c}, V. \& Riechers, D.~A.\ 2011, \apj, 730, 64

\bibitem[Soifer et al.(2000)]{Soifer2000} Soifer, B.~T., Neugebauer, G., Matthews, K., et al.\ 2000, \aj, 119, 509


\bibitem[Solomon \& Vanden Bout(2005)]{Solomon2005} Solomon, P.~M. \& Vanden Bout, P.~A.\ 2005, \araa, 43, 677

\bibitem[Speranza et al.(2022)]{Speranza2022} Speranza, G., Ramos Almeida, C., Acosta-Pulido, J.~A., et al.\ 2022, arXiv:2206.15347

\bibitem[Spoon \& Holt(2009a)]{Spoon2009a} Spoon, H.~W.~W. \& Holt, J.\ 2009a, \apjl, 702, L42

\bibitem[Spoon et al.(2009b)]{Spoon2009b} Spoon, H.~W.~W., Armus, L., Marshall, J.~A., et al.\ 2009b, \apj, 693, 1223

\bibitem[Stanghellini et al.(2005)]{Stanghellini2005} Stanghellini, C., O'Dea, C.~P., Dallacasa, D., et al.\ 2005, \aap, 443, 891

\bibitem[Tadhunter et al.(2011)]{Tadhunter2011} Tadhunter, C., Holt, J., Gonz{\'a}lez Delgado, R., et al.\ 2011, \mnras, 412, 960

\bibitem[Tadhunter et al.(2014)]{Tadhunter2014} Tadhunter, C., Morganti, R., Rose, M., et al.\ 2014, \nat, 511, 440

\bibitem[Tadhunter et al.(2018)]{Tadhunter2018} Tadhunter, C., Rodr{\'\i}guez Zaur{\'\i}n, J., Rose, M., et al.\ 2018, \mnras, 478, 1558

\bibitem[Togi \& Smith(2016)]{Togi2016} Togi, A. \& Smith, J.~D.~T.\ 2016, \apj, 830, 18

\bibitem[U et al.(2012)]{U2012} U, V., Sanders, D.~B., Mazzarella, J.~M., et al.\ 2012, \apjs, 203, 9

 \bibitem[\protect\citeauthoryear{Vacca}{2003}]{Vacca2003} 
Vacca, W.D., Cushing, M.C.,  \&  Rayner, J. T.,  2003, PASP, 115, 389

\bibitem[Veilleux et al.(1997)]{Veilleux1997} Veilleux, S., Sanders, D.~B., \& Kim, D.-C.\ 1997, \apj, 484, 92

\bibitem[Veilleux et al.(2009)]{Veilleux2009} Veilleux, S., Rupke, D.~S.~N., Kim, D.-C., et al.\ 2009, \apjs, 182, 628

\bibitem[Villar-Mart{\'\i}n et al.(2013)]{Villar2013} Villar-Mart{\'\i}n, M., Rodr{\'\i}guez, M., Drouart, G., et al.\ 2013, \mnras, 434, 978

\bibitem[Villar Mart{\'\i}n et al.(2015)]{Villar2015} Villar Mart{\'\i}n, M., Bellocchi, E., Stern, J., et al.\ 2015, \mnras, 454, 439.

\bibitem[Villar-Mart{\'\i}n et al.(2017)]{Villar2017} Villar-Mart{\'\i}n, M., Emonts, B., Cabrera Lavers, A., et al.\ 2017, \mnras, 472, 4659

\bibitem[Villar Mart{\'\i}n et al.(2020)]{Villar2020} Villar Mart{\'\i}n, M., Perna, M., Humphrey, A., et al.\ 2020, \aap, 634, A116

\bibitem[Villar Mart{\'\i}n et al.(2021)]{Villar2021} Villar Mart{\'\i}n, M., Emonts, B.~H.~C., Cabrera Lavers, A., et al.\ 2021, \aap, 650, A84.

\bibitem [\protect\citeauthoryear{Wright}{2006}]{Wright2006}
Wright, E.L. 2006 PASP, 118, 1711

\bibitem[Zakamska et al.(2003)]{Zakamska2003} Zakamska, N.~L., Strauss, M.~A., Krolik, J.~H., et al.\ 2003, \aj, 126, 2125

\bibitem[Zakamska(2010)]{Zakamska2010} Zakamska, N.~L.\ 2010, \nat, 465, 60

\bibitem[Zhu et al.(2017)]{Zhu2017} Zhu, H., Tian, W., Li, A., et al.\ 2017, \mnras, 471, 3494

\end{thebibliography}
\end{document}